\documentclass[amsmath,amssymb,aps,prx,reprint,superscriptaddress]{revtex4-2}

\usepackage{amsmath}
\usepackage{xcolor}
\usepackage{amsfonts,amssymb}
\usepackage{graphicx}
\usepackage{hyperref}
\usepackage{academicons}
\usepackage{orcidlink}

\begin{document}

\title{Modeling intercalation chemistry with multi-redox reactions by sparse lattice models in disordered rocksalt cathodes}

\author{\text{Peichen Zhong}}
\thanks{These two authors contributed equally.}
\affiliation{Department of Materials Science and Engineering, University of California, Berkeley, California 94720, United States}
\affiliation{Materials Sciences Division, Lawrence Berkeley National Laboratory, California 94720, United States}

\author{\text{Fengyu Xie}}
\thanks{These two authors contributed equally.}
\affiliation{Department of Materials Science and Engineering, University of California, Berkeley, California 94720, United States}
\affiliation{Materials Sciences Division, Lawrence Berkeley National Laboratory, California 94720, United States}

\author{\text{Luis Barroso-Luque}}
\affiliation{Department of Materials Science and Engineering, University of California, Berkeley, California 94720, United States}
\affiliation{Materials Sciences Division, Lawrence Berkeley National Laboratory, California 94720, United States}

\author{\text{Liliang Huang}}
\affiliation{Materials Sciences Division, Lawrence Berkeley National Laboratory, California 94720, United States}

\author{\text{Gerbrand Ceder}}
\email[]{gceder@berkeley.edu}
\affiliation{Department of Materials Science and Engineering, University of California, Berkeley, California 94720, United States}
\affiliation{Materials Sciences Division, Lawrence Berkeley National Laboratory, California 94720, United States}

\date{\today}

\begin{abstract}

Modern battery materials can contain many elements with substantial site disorder, and their configurational state has been shown to be critical for their performance. The intercalation voltage profile is a critical parameter to evaluate the performance of energy storage. The application of commonly used cluster expansion techniques to model the intercalation thermodynamics of such systems from \textit{ab-initio} is challenged by the combinatorial increase in configurational degrees of freedom as the number of species grows. Such challenges necessitate efficient generation of lattice models without over-fitting and proper sampling of the configurational space under charge balance in ionic systems. In this work, we introduce a combined approach that addresses these challenges by (1) constructing a robust cluster-expansion Hamiltonian using the sparse regression technique, including $\ell_0\ell_2$-norm regularization and structural hierarchy; and (2) implementing semigrand-canonical Monte Carlo to sample charge-balanced ionic configurations using the table-exchange method and an ensemble-average approach. These techniques are applied to a disordered rocksalt oxyfluoride Li$_{1.3-x}$Mn$_{0.4}$Nb$_{0.3}$O$_{1.6}$F$_{0.4}$ (LMNOF) which is part of a family of promising earth-abundant cathode materials. The simulated voltage profile is found to be in good agreement with experimental data and particularly provides a clear demonstration of the Mn and oxygen contribution to the redox potential as a function of Li content. 
\end{abstract}


\maketitle

\section{Introduction}
The market for electric-vehicle high-energy-density Li-ion batteries has witnessed tremendous growth over the past decade \cite{Olivetti2017,Goodenough2010}. The increasing demand for electrical energy storage requires further development of the high-energy-density-based cathode materials in rechargeable Li-ion batteries. Current cathode materials are mostly limited to layered Li(Ni,Mn,Co)O$_2$ (NMC) variants. The high cost of Ni and Co limits the large-scale expansion of Li-ion batteries with NMC-type cathodes. Recently developed disordered rocksalt with Li-excess (DRX) are promising earth-abundant cathode materials, which can enable scaling of Li-ion energy storage to several TWh/year production. Although these compounds have no long-range cation order, the interactions between species generate short-range ordering (SRO), which critically affects the electrochemical performance \cite{Ji2019_NatComm_SRO}. The broad chemical flexibility and the wide variety of chemical environments that can be created for Li and the transition metals (TMs) that can be used by SRO and chemistry provide new opportunities to improve the cathode performance. Examples include improving cyclability via fluorine/vacancy doping of the anion sublattice \cite{Richards2018_fluorination, Huang2022_oxyvac}, enhancing the rate capability by engineering the cation SRO \cite{Ji2019_NatComm_SRO}, and achieving zero-strain cathodes for solid-state batteries \cite{zhang2022high_entropy_NMC, Zhou2022_zero_strain, Zhao2022_zerostrain}.

Over the last two decades, many properties of Li-ion battery materials have been successfully predicted by first-principles calculations \cite{Meng2009_review, Urban2016npj, VanDerVen2020}. The equilibrium voltage is one of the fundamental quantities that help characterize the electrochemical performance of a particular material and is defined by the difference in Li chemical potentials between the cathode and anode:
\begin{equation}
    V = - \frac{\mu_{\text{Li}}^{\text{cathode}} - \mu_{\text{Li}}^{\text{anode}}}{zF}.
    \label{eq:voltage_mu}
\end{equation}
In Eq. \eqref{eq:voltage_mu}, $z$ is the charge transferred per ion, $F$ is Faraday's constant, and $\mu_{\text{Li}}$ is the chemical potential of Li. For example, considering a Li transition-metal oxide Li$_x$TMO$_2$ ($x_1\leq x \leq  x_2$) as the cathode and Li metal as the anode with the cell reaction
\begin{equation}
    \text{Li}_{x_1}\text{TMO}_2  \longrightarrow \text{Li}_{x_2}\text{TMO}_2 + (x_1-x_2)\text{Li},
    \label{eq:reaction}
\end{equation}
the approximated equilibrium voltage can be computed as \cite{aydinol1997_abinit_voltage}
\begin{equation}
\Bar{V}(x_1, x_2) \approx -\frac{E_{\text{Li}_{x_1}\text{TM}\text{O}_2} - E_{\text{Li}_{x_2}\text{TM}\text{O}_2} - (x_1 - x_2)E_{\text{Li}}}{F(x_1 - x_2)}.
\label{eq:avg_voltage}
\end{equation}
The internal energy of the bcc Li metal $E_{\text{Li}}$, the lithiated structure $E_{\text{Li}_{x_1}\text{M}\text{O}_2}$, and the delithiated structure ($E_{\text{Li}_{x_2}\text{M}\text{O}_2}$) can be obtained from first-principles density functional theory (DFT). In this approach, the entropic effect is assumed to be small at low temperatures, and the change in internal energy is used to approximate the chemical potential change. By computing the formation energy of Li$_x$TMO$_2$ structures with varied Li concentrations $x$, a convex hull can be constructed from the energy of ground-state structures at each concentration. A piece-wise voltage profile can then be built from Eq. \eqref{eq:avg_voltage} via the ground states on the convex hull by using relevant constructive values of $x$ on the hull.

\begin{figure*}[t]
\centering
\includegraphics[width=\linewidth]{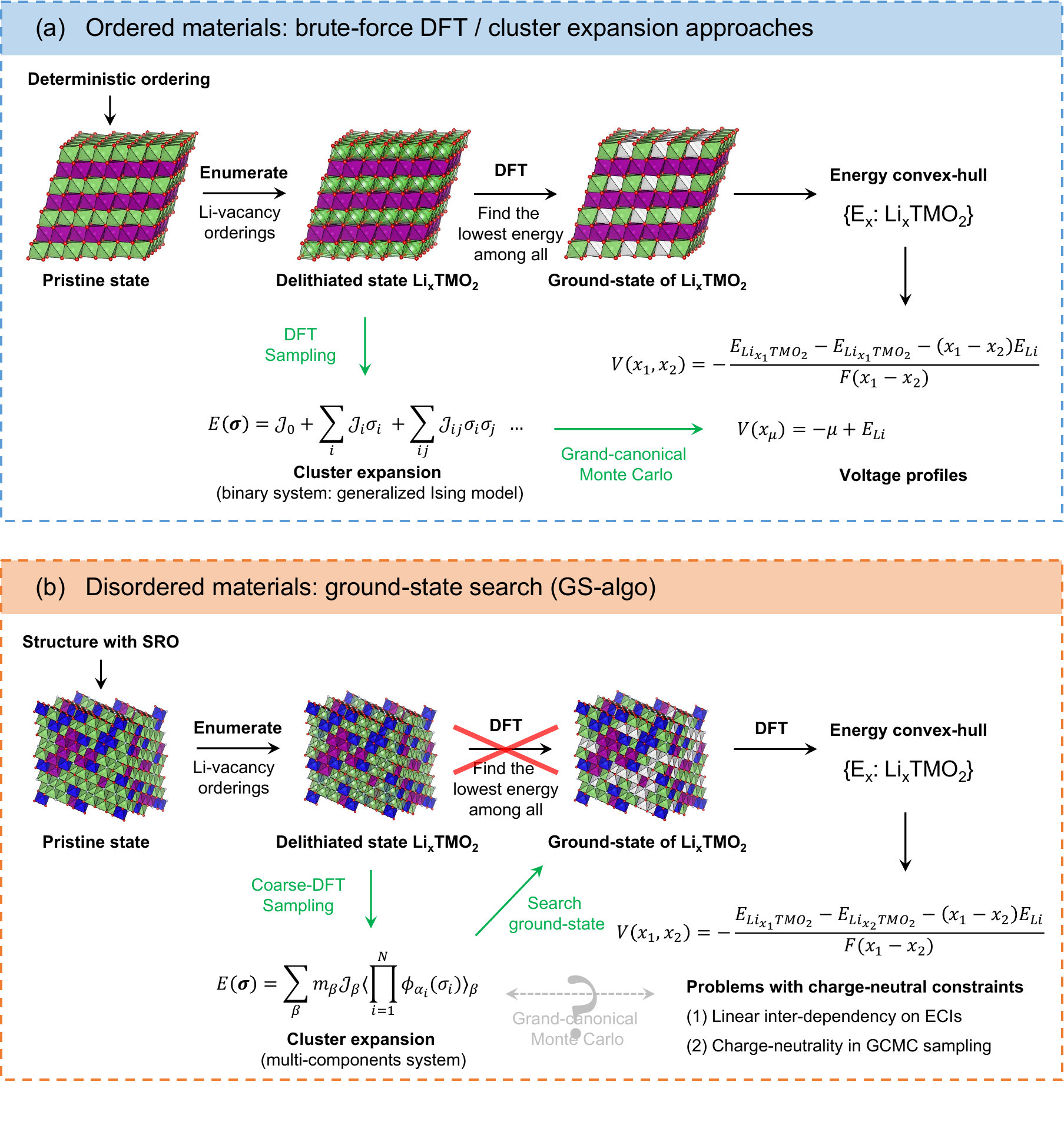}
\caption{An overview of reported methods for computing voltage profiles in (a) ordered and (b) disordered electrode materials. The green arrows represent the configurational samplings that can be accelerated by using cluster expansion as an effective Hamiltonian fitted from DFT calculations.}
\label{fig:previous_algo}
\end{figure*}

Unlike simple Li transition-metal oxides (LiTMO$_2$), a typical DRX cathode (Li$_{1+x}$M'$_a$M''$_b$O$_{2-y}$F$_y$) is composed of three major components: (1) the redox-active species M', which provides electron redox; (2) the inert high-valent TM M'', which charge compensates for the Li excess and stabilizes disordered structures \cite{Urban2017_PRL}; (3) fluorine, which can improve the cyclability and allows more Li excess to be accommodated without losing TM redox by lowering the overall anion valence \cite{Lun2019AEM_F_cycle}. When the cathode is charged/discharged, Li is removed/inserted into the cathode structure accompanied by oxidation/reduction, resulting in various oxidation states among the redox-active metal M' and oxygen atoms. Different oxidation states of a TM can exhibit very different local chemistry preferences (e.g., Mn$^{3+}$ has a substantial Jahn--Teller effect compared to Mn$^{2+/4+}$). To capture these chemical differences in simulations, different valence states of the same elements must be treated as distinct species. This treatment is called \emph{charge decoration}, which has been demonstrated to be essential in capturing the electronic entropy effect to construct the correct phase diagram in some compounds \cite{Zhou2006electronic_entropy}. Charge decoration intrinsically increases the number of components, and therefore the complexity of modeling the intercalation voltage profiles of DRXs.

\subsection{Intractability of composition enumeration}
To obtain the voltage profile of a DRX material, most previous studies have used the convex-hull construction approach by finding the ground states (GS), which we refer to as the \emph{GS-algo}. In this approach, one tries to find the low-energy structures at varied Li content $x$ using a variety of algorithms \cite{Lee2021_critivity, Li2021_NatureChem}. These low-energy configurations are calculated by DFT to construct the piece-wise voltage profile following Eq. \eqref{eq:avg_voltage}. The GS-algo has shown reasonable predictions for voltages and redox mechanisms \cite{Yao2018_LiMnO, li2018LiCo_NonEqui, Ji2019_Ni, Li2021_NatureChem, McColl2022_NatComm, Lee2021_critivity}. When a high number of components and valence states are present, the GS-algo can become impractical as all possible valence combinations at each stage of delithiation must be enumerated. For example, when evaluating a delithiated supercell of composition Li$_{21}\square_{18}$Mn$_{12}$Nb$_{9}$O$_{48}$F$_{12}$, the combination of valence in Mn and O can take Mn$^{3+}_{6}$Mn$^{4+}_{6}$, Mn$^{2+}_{1}$Mn$^{3+}_{4}$Mn$^{4+}_{7}$, Mn$^{2+}_{2}$Mn$^{3+}_{2}$Mn$^{4+}_{8}$, and even Mn$^{3+}_{7}$Mn$^{4+}_{5}$O$^{-}_1$, etc. Enumerating all the possible compositions and searching for the possible ground states under each charge-decorated composition are NP-hard problems and become intractable, in particular when the supercell grows. To resolve the enumeration problem, Monte Carlo (MC) sampling is a better choice for studying configurational energetics in a high-dimensional space.

\subsection{Fast growth of cluster basis caused by charge decoration}
To bridge the gap between 0K ground states and sampling at finite temperatures, MC simulation with a cluster expansion (CE) as an effective Hamiltonian is typically used for intercalation chemistry in ordered cathode materials \cite{aangqvist2019icet,ShiSiQi2016_CE_review, van2002_ATAT}. The CE casts the energy as a function of the occupancy of atoms on a set of predefined sites. For example, in a multicomponent system, the energy is expanded as 
\begin{equation}
	E(\boldsymbol{\boldsymbol{\sigma}}) = \sum_{\beta} m_{\beta} J_{\beta}\left\langle \Phi_{\boldsymbol{\alpha} \in \beta} \right\rangle_{ \beta} + \frac{E_0}{\varepsilon_r},~ \Phi_{\boldsymbol{\alpha}} = \prod_{i=1}^N \phi_{\alpha_i}(\sigma_i).
	\label{eq:ce_definition}
 \end{equation}
A configuration $\boldsymbol{\sigma}$ represents a specific occupancy state of species on all the system sites, where $\sigma_i$ describes which species sits on the $i$-th site of the lattice.  The cluster basis function $\Phi_{\boldsymbol{\alpha}} = \prod_{i=1}^N \phi_{\alpha_i}(\sigma_i)$ are the product of site basis functions $\phi_{\alpha_i}(\sigma_i)$ across a collection $\alpha$ of multiple sites. They are taken as the average over the crystal symmetry orbits $\beta$, forming a complete basis to expand the scalar energy function on the configuration space. The expansion coefficients $J_{\beta}$ are called effective cluster interactions (ECIs). The electrostatic energy (Ewald energy $E_0/\varepsilon_r$) is also included to capture long-range electrostatic interactions \cite{Seko2014_longrange} ($E_0$ is the unscreened electrostatic energy, and $1/\varepsilon_r$ is fitted as one of the ECIs ($1/\varepsilon_r\geq 0$). We refer readers to Ref. \cite{Barroso-Luque2022_theory} for a more comprehensive description of the CE formalism in ionic systems. 

\begin{figure}[h]
\centering
\includegraphics[width=\linewidth]{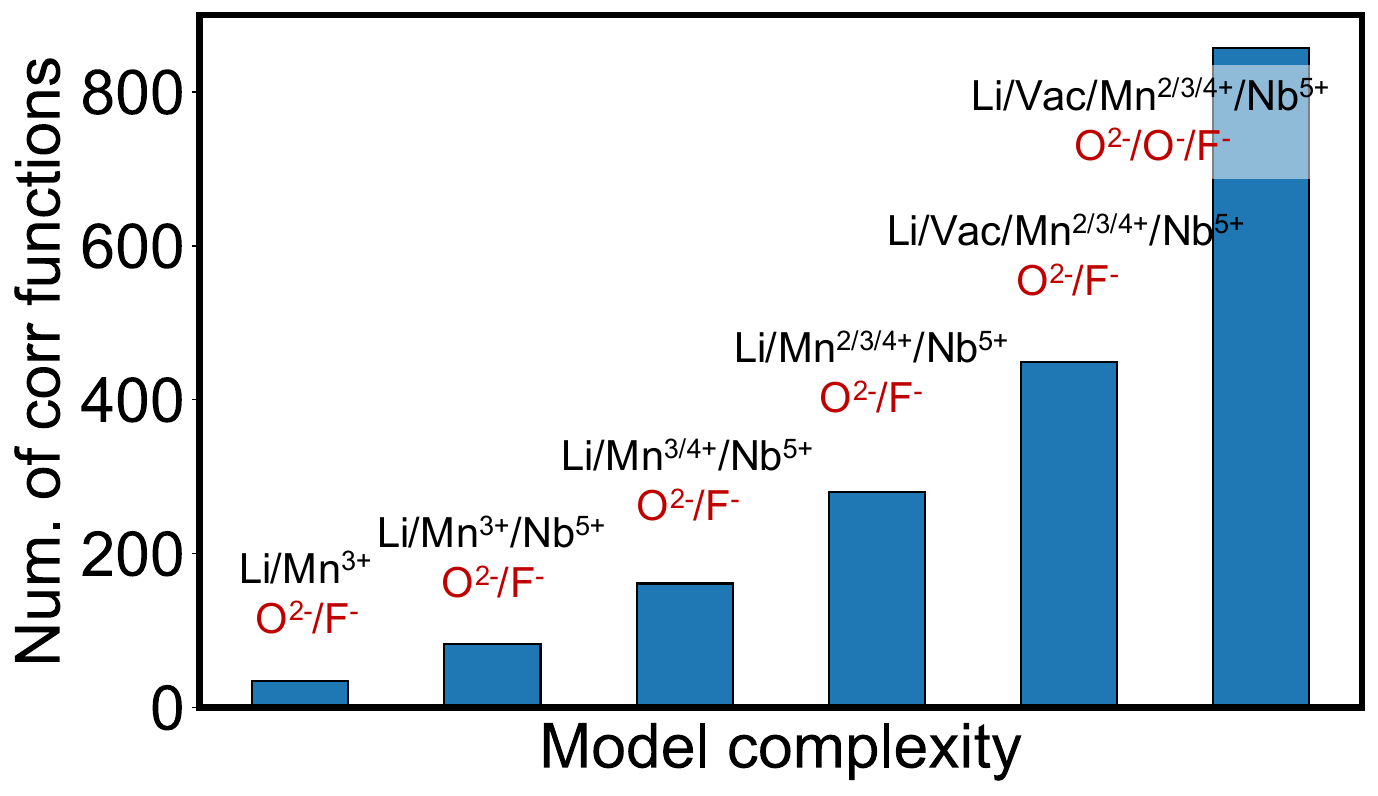}
\caption{An illustration of cluster basis growth: number of correlation functions  vs. number of components included in the CE with pair/triplet/quadruplet cutoff radius of 7/4/4 \AA\ based on a rocksalt primitive cell with lattice parameter $a=3$ \AA.}
\label{fig:basis_explosion} 
\end{figure}

Since the cluster site basis functions are defined by the number of components \cite{VandeWalle2009}, the charge decoration can significantly increase the model complexity of the CE, resulting in a rapid growth in the number of clusters. Figure \ref{fig:basis_explosion} illustrates how the number of cluster basis functions grows as the number of cation species included in the DRX increases, where the cutoff radius is fixed with pair/triplet/quadruplet interactions up to 7/4/4 \AA. For the full description of Mn and O redox, more than 800 ECIs are predefined and need to be fitted, whereas the number of DFT calculations is typically much smaller than that of predefined ECIs. When building a CE, the ECIs $\boldsymbol{J}^* = \{J_\beta\}$ can be obtained by fitting the DFT energy $\boldsymbol{E}_{\text{DFT},S}$ of training structures $S$ to their correlation functions with (regularized) linear regression:
\begin{equation}
    \boldsymbol{J}^* = \arg\min_{\boldsymbol{J}} ~ ||\boldsymbol{E}_{\text{DFT},S} -\boldsymbol{\Pi}_S\boldsymbol{J}||^2_2 + \rho(\boldsymbol{J}),
    \label{eq:linear_regression}
\end{equation}
where $\boldsymbol{\Pi}_S$ is the feature matrix formed by the correlation functions and $\rho(\boldsymbol{J})$ is a regularization term. 
High-component CE can easily be overfitted as the rank of the feature matrix ($\boldsymbol{\Pi}_S$) is typically smaller than the dimension of ECIs dim($\boldsymbol{J}$). The low-rank structure (referred to as \emph{rank deficiency}) of the feature matrix $\boldsymbol{\Pi}_S$ in Eq. \eqref{eq:linear_regression} requires selecting the most physically informative ECIs and avoiding overfitting in CE fitting \cite{Nelson2013, Seko_sparseL1}. The rank deficiency can also arise from other physical constraints in ionic systems. For example, charge balance creates a linear dependency between the number of charge-decorated species and the corresponding correlation functions. And another challenge is that the training structures calculated with DFT are predominantly low in energy, and such low-energy structures often narrow the configurational space that can be represented in the training structures, which in principle, high-energy configurations could be included. DFT tends not to cover such configurations or relax them to lower-energy configurations by moving ions and electrons. The number of non-zero ECIs must be constrained ($||\boldsymbol{J}||_0 \leq \text{rank}(\boldsymbol{\Pi}_S)$) to prevent ECI fitting from being an underdetermined problem. This constraint can be achieved by properly introducing $\mu||\boldsymbol{J}||_0$ as a regularization term in Eq. \eqref{eq:linear_regression} to penalize the number of non-zero ECIs and impose sparsity \cite{Zhong2022L0L2}.

\subsection{Charge-neutrality constraint in MC sampling}
After fitting the ECIs, MC simulations can be used to sample the energy of configurations under finite temperatures. Applying semigrand-canonical Monte Carlo (sGCMC) sampling is most suited for calculating voltage profiles \cite{Puchala2023_CASM, VanderVen1998_LiCoO2}. The relation between the Li content $x$ and a Li chemical potential can be obtained by applying the Metropolis--Hastings algorithm with the Boltzmann distribution
\begin{equation}
    f(E(\boldsymbol{\sigma}),\mu) \propto \exp \left(- \frac{E(\boldsymbol{\sigma})-\mu_{\text{Li}}\cdot x_{\text{Li}} N}{k_BT} \right).
    \label{eq:Boltzmann}
\end{equation}
In Eq. \eqref{eq:Boltzmann}, $E$ is the energy of the configuration $\boldsymbol{\sigma}$ given by the CE Hamiltonian, $\mu_{\text{Li}}$ is the Li chemical potential, $x_{\text{Li}}$ is the Li content in the configuration $\boldsymbol{\sigma}$, $N$ is the total number of Li and vacancy sites, $k_B$ is the Boltzmann constant, and $T$ is the simulation temperature.  Computing voltage profiles using sGCMC has been successful in the study of several simple binary electrode materials, such as LiCoO$_2$/LiNiO2 \cite{VanderVen1998_LiCoO2, Wolverton1998_LiCO2,houchins2020_MLP_gcmc}, MgTiS$_2$ \cite{Kolli2018_MgTiS2}, and Li$_{3}$V$_2$O$_5$ \cite{guo2022intercalation}.

When charge-decorated CEs are used, the requirement of charge neutrality must be enforced in MC sampling. Because the training set only includes charge-balanced structures with no information about the charge-unbalanced structures, the energy predicted by the CE will be unphysical if configurations with non-zero net charge are assessed during sGCMC. Techniques for enforcing strict charge neutrality in sGCMC have been applied to several electrolyte systems \cite{valleau1980_chg_GCMC, Deng2020_Na_electrolyte}; however, few have been demonstrated in a system with complex redox reactions such as DRX \cite{xie2022grand}.

To overcome all the abovementioned issues, we propose a voltage-calculation framework that combines several state-of-the-art methods in CE-MC. With this framework, we demonstrate how to correctly model the intercalation voltage profile in DRX and, more generally, any complex ionic systems with redox-active ions and configurational disorder. In the following methodology sections, we will introduce the construction of a robust and predictive cluster-expansion Hamiltonian with $\ell_0\ell_2$-norm regularized sparse regression \cite{Zhong2022L0L2}, demonstrate an effective sampling strategy of the intercalation stages with sGCMC under charge balance using the table-exchange (TE) method \cite{xie2022grand}, and illustrate an ensemble average method over representative structures to handle various chemical environments.
In the results section, the equilibrium voltage profile of Li$_{1.3-x}$Mn$_{0.4}$Nb$_{0.3}$O$_{1.6}$F$_{0.4}$ (LMNOF) is presented and compared with the experiments. To explain the redox mechanism, we analyze the proportion of multiple redox-active species at varied Li content. We find that the calculated voltage profile and redox mechanism agree well with experiments and argue that the ability of our method to describe oxygen redox in the Li-excess cathode accurately is particularly noticeable.

\section{Methodology}
\subsection{Training structure generation for DFT}

To describe DRX materials well, two configurational degrees of freedom need to be accurately represented: the Li/vacancy interactions and the different local chemical environments (i.e., the SRO of TMs and anions). We propose the following two-step procedure for generating the training set:

(1) \emph{Pristine states}: Generate several fully lithiated structures with different transition-metal and anion configurations in relatively small supercells (e.g., supercell structures with $10 \times$ or $20 \times$ the formula unit of Li$_{1.3}$Mn$_{0.4}$Nb$_{0.3}$O$_{1.6}$F$_{0.4}$). These structures can be generated from canonical MC samplings using a pre-fitted cluster expansion as the effective Hamiltonian or using solely the electrostatic energy for simplicity.

(2) \emph{Delithiated states}: Starting from the structures generated in Step 1, fix the TM and O/F orderings and enumerate different Li/vacancy configurations at varied Li contents (e.g., $x = 0.3/0.5/0.7$ in Li$_{1.3-x}$). As the total number of enumerated structures can be large, one can further sort the structures at each Li content by their electrostatic energy and only keep the low-energy ones. 

All the sampled structures are in the chemical space of Li$_{1.3-x}$Mn$_{0.4}$Nb$_{0.3}$O$_{1.6}$F$_{0.4}$, and this two-step procedure covers different Li/vacancy orderings in varied local chemical environments formed by TM and anion SRO to be calculated with DFT.

\subsection{Sparse regression for charge-decorated CE}
\begin{figure*}[t]
\centering
\includegraphics[width=\linewidth]{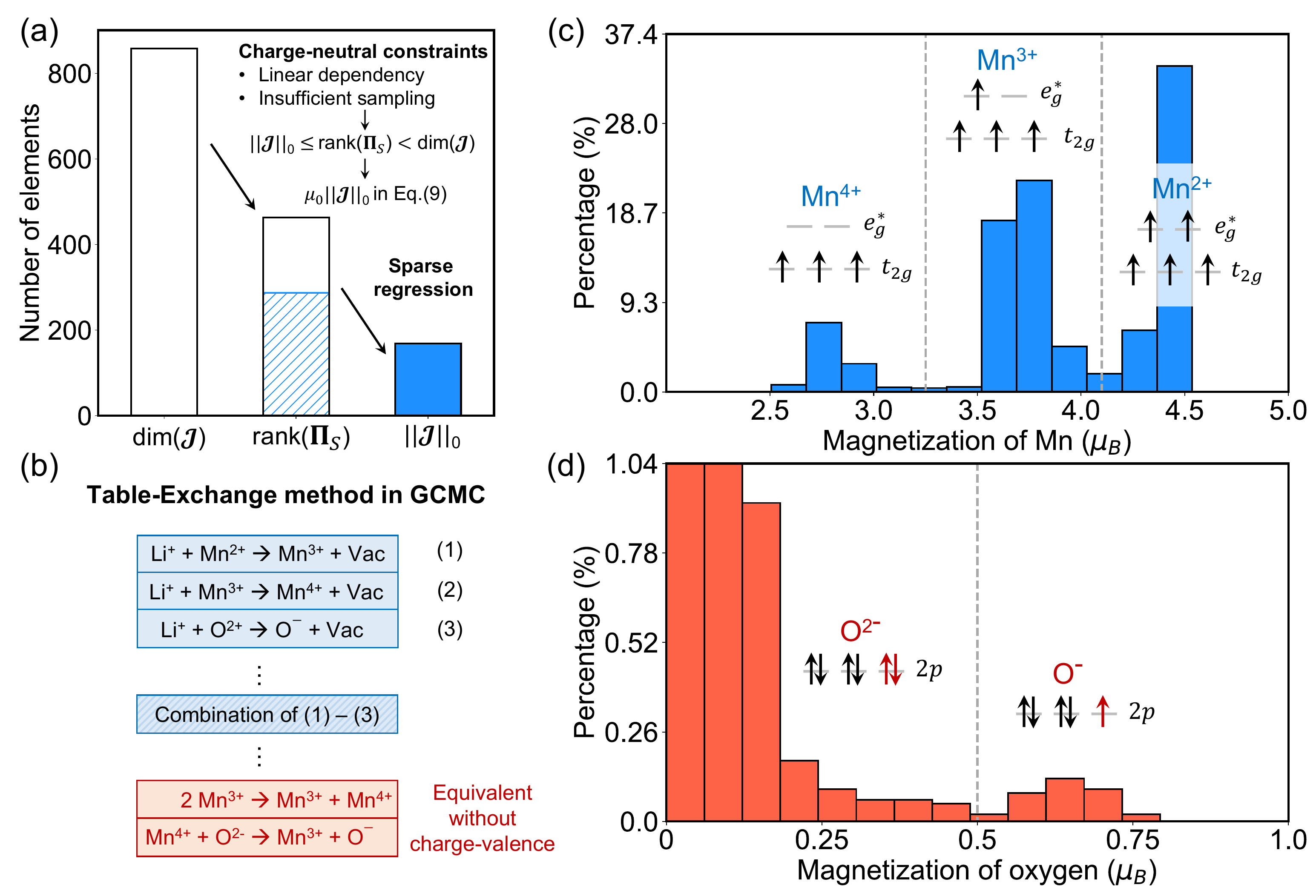}
\caption{(a) An illustration of the rank deficiency in the ECI fitting.  The left empty bar represents the dimension of predefined ECIs with dim($\boldsymbol{J}$) = 858. The middle empty bar indicates the number of structures sampled in the feature matrix with dim$(\boldsymbol{\Pi}_S) = (463\times 858)$. The blue shaded area overlapping the middle bar indicates a feature matrix of rank$(\boldsymbol{\Pi}_S) = 287$. The right solid blue bar represents the number of non-zero ECIs after an $\ell_0\ell_2$-norm regularized sparse regression, giving $||\boldsymbol{J}||_0 = 169$. (b) An illustration of the TEs used in charge-balanced sGCMC.
(c)--(d) The distribution of total on-site magnetizations of Mn (c) and O (d) atoms among all DFT-r$^2$SCAN calculated structures in the chemical space of Li$_{1.3-x}$Mn$_{0.4}$Nb$_{0.3}$O$_{1.6}$F$_{0.4}$. The valence of each Mn and O atom is determined by the on-site magnetization. From the histogram, the classification boundary between Mn$^{4+/3+}$ and Mn$^{3+/2+}$ is estimated to be $3.25 \mu_B$ and $4.1 \mu_B$, and the O$^{-}$ classification is estimated to be $0.5\mu_B$. (As the percentage of O$^{2-}$ with a low magnetization is too large compared to the percentage of O$^-$, the panel (d) is truncated on the y-axis.) }
\label{fig:tech_imrove}
\end{figure*}

To obtain effective valence states of redox-active Mn and O species from a DFT calculated configuration, the on-site magnetization can be used \cite{Barroso-Luque2022_theory, Yang2022_npj}. For example, Figure \ref{fig:tech_imrove}(c) and (d) show the distribution of magnetic moments representing Mn$^{2+/3+/4+}$ and O$^{2-/-}$ in our set of 463 DFT-calculated structures. The valence of each Mn and O atom is classified by the site magnetization using 3.25$\mu_B$ for distinguishing Mn$^{4+/3+}$, 4.1$\mu_B$ for separating Mn$^{3+/2+}$, and 0.5$\mu_B$ to indicate O$^{-}$  \cite{Seo2016_NatChem} ($\mu_B$ is the Bohr magneton). For the sparse regression of ECIs, we apply the $\ell_0\ell_2$-norm regularization with hierarchy constraints \cite{Zhong2022L0L2}. The ECIs are optimized in the following mixed-integer quadratic programming problem:
\begin{align}
&\min_{\boldsymbol{J}} ~ \boldsymbol{J}^T\boldsymbol{\Pi}_S^T\boldsymbol{\Pi}_S\boldsymbol{J}^T - 2\boldsymbol{E}_{\text{DFT}}^T\boldsymbol{\Pi}_{S}\boldsymbol{J} + \mu_0\sum_{ c\in \boldsymbol{C}} z_{0,c} +\mu_2||\boldsymbol{J}||_2^2 \label{eq:corrL0L2}\\
\nonumber
&\begin{array}{r@{~}l@{}l@{\quad}l}
\boldsymbol{s.t.} \quad Mz_{0,c} &\geq J_c, ~\forall c\in \boldsymbol{C}\\
Mz_{0,c} &\geq -J_c,~\forall c\in \boldsymbol{C}\\
z_{0,b} &\leq z_{0,a},  ~\forall a \subset b, ~ \{a,b\} \in \boldsymbol{C}\\
z_{0,c} &\in \{0,1\}, ~\forall c\in \boldsymbol{C},
\end{array}
\end{align}
where $\boldsymbol{\Pi}_S$ is the feature matrix, $\boldsymbol{J}$ are the ECIs, $z_{0,c}$ is the slack variable representing $J_c = 0, z_{0,c} = 0$, and $ J_c \neq 0,  z_{0,c} = 1$. $M = 100$ is set to constrain the optimization boundaries, and  $||\boldsymbol{J}||_2^2 = \boldsymbol{J}^T\boldsymbol{J}$ is a ridge regression term ($\ell_2$-norm). We refer the readers to Ref. \cite{Zhong2022L0L2} for a detailed description of this approach which we found to lead to relatively sparse but accurate CE. 

The CE Hamiltonian was constructed with pair interactions up to 7 \AA, triplet interactions up to 4 \AA, and quadruplet interactions up to 4 \AA\ based on a rocksalt primitive cell with lattice parameter $a=3$ \AA\ leading to a possible 858 ECIs (including the constant term $J_0$). The ECIs were fitted using 463 training structures, forming a feature matrix of rank$(\boldsymbol{\Pi}_S) = 287$. The resulting ECIs using the sparse regression in Eq. \eqref{eq:corrL0L2} contain 169 non-zero elements ($||\boldsymbol{J}||_0 = 169$). The relationship between dimension, rank and the number of non-zero elements is illustrated in Fig. \ref{fig:tech_imrove}(a).

\subsection{Charge-balanced Monte Carlo sampling}
\begin{figure*}[t]
\centering
\includegraphics[width=\linewidth]{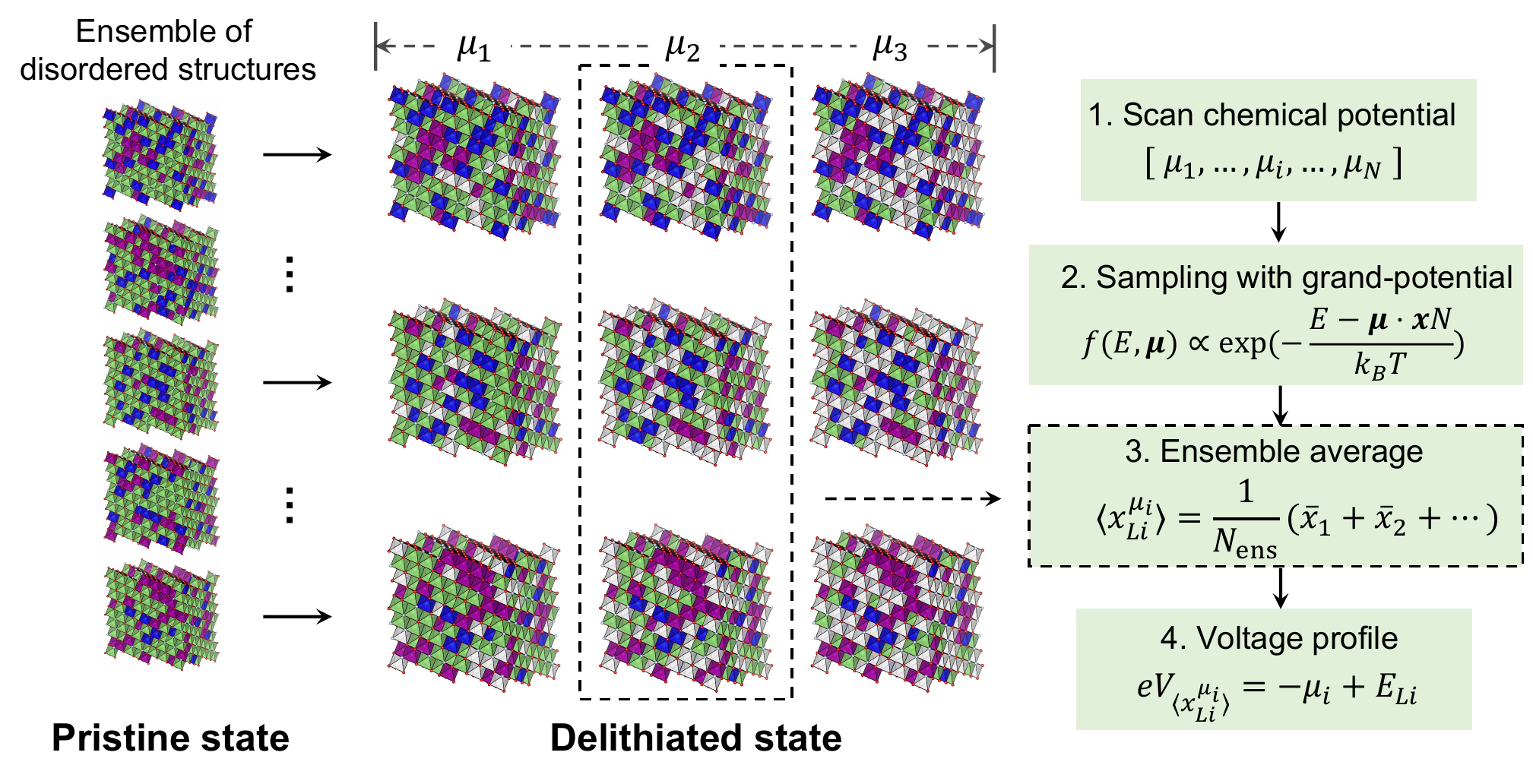}
\caption{An illustration of charge-balanced sGCMC on an ensemble of disordered structures (Green: Li, white: vacancy, other colors: different TM species, anions are not displayed). An ensemble of fully lithiated disordered structures is first generated and referred to as the pristine state. The sGCMC is performed to topotactically delithiate each pristine structure at decreasing Li chemical potentials $\mu_{\text{Li}}$ (i.e., increasing voltages $V$). The content of species is averaged over each sGCMC sampling and all the pristine structures.}
\label{fig:GCMC_flow}
\end{figure*}

We applied sGCMC simulations on the Li/vacancy occupancy and the charge decoration degrees of freedom to obtain the voltage curve of Li$_{1.3-x}$Mn$_{0.4}$Nb$_{0.3}$O$_{1.6}$F$_{0.4}$ composition.  The delithiation is achieved by step-wise removal of Li atoms. In each MC step, the Li$^+$ is removed/inserted, accompanied by the oxidation/reduction of an Mn or O atom. The net charge of each configuration is maintained at zero by only executing a combination of site occupancy changes that are charge neutral.  This type of MC step is referred to as table exchange (TE) \cite{xie2022grand,valleau1980_chg_GCMC}. In our calculations, we used the following three elemental classes of perturbations:
\begin{enumerate}
\centering
    \item Li$^+$ + Mn$^{2+} \rightarrow$  Mn$^{3+}$ + Vac. 
    \item Li$^+$ + Mn$^{3+} \rightarrow$  Mn$^{4+}$ + Vac. 
    \item Li$^+$ + O$^{2-} \rightarrow$  O$^{-}$ + Vac.
\end{enumerate} 
Any other charge-conserving MC step can be expressed as a linear combination of these three classes and their inverses. For example, charge transfer such as $2\text{Mn}^{3+} \rightarrow \text{Mn}^{2+} + \text{Mn}^{4+}$ and $\text{Mn}^{4+} + \text{O}^{2-} \rightarrow \text{Mn}^{3+} + \text{O}^{-} $ can be achieved by a combination of elementary perturbations. As the intercalation is assumed to be topotactic, Mn, O, and F ions do not change sites. The acceptance probabilities of each MC step are scaled to ensure detailed balance (see Appendix). 

In a series of sGCMC simulations, the chemical potentials are scanned between two limiting values $\mu_{\text{Li}} \in [\mu_{\min}, \mu_{\max}] $ at finite temperature. For each sGCMC simulation with a given Li chemical potential $\mu$, the content of each charge-decorated species is averaged over the sampled MC structures from the equilibrium. The Li chemical potential can be converted into the cathode voltage using $V = -(\mu - E_{\text{Li}}) / e$. 

In this work, the numerical simulations were implemented with the open-source software package \texttt{smol} \cite{Barroso-Luque2022smol}.

\subsection{The ensemble average method}
Unlike ordered intercalation compounds which are characterized by a small number of local environments, the configurational disorder in DRX creates an abundance of local chemical environments, leading to a significant variation of Li extraction energy from different sites \cite{Abdellahi2016, Squires2023_SRO}. This multitude of environments is difficult to capture in a single small unit cell. Instead, we sample the delithiation from multiple distinct structures and obtain the true lithiation curve as the ensemble average of them.

We use a canonical MC simulation to generate an ensemble of fully lithiated (i.e., pristine) configurations $\{ \boldsymbol{\sigma}_i\}$, from which multiple representative structures are drawn. The CE model used to generate the canonical MC structures is reported in Ref. \cite{cationVacancy_liliang}. To recover the actual SRO in DRX, we require that the average of physical quantities in the collection of representative structures be approximately equal to the actual ensemble average, assuming infinitely many structures are drawn from the ensemble ($\langle\boldsymbol{\Pi}\rangle_{\text{ens}} \approx \langle\boldsymbol{\Pi}\rangle_{\infty} = \Bar{\boldsymbol{\Pi}}_{\text{SRO}}$). For simplicity, in this work, we verify that the average energy  $\langle E\rangle_{\text{ens}} = \langle\boldsymbol{\Pi}\rangle_{\text{ens}} \cdot \boldsymbol{J} \approx \Bar{\boldsymbol{\Pi}}_{\text{SRO}} \cdot \boldsymbol{J}$ converges to the ensemble average with an increasing number of selected structures. 

Each simulated fully lithiated structure contains 120 atoms (Li$_{39}$Mn$_{12}$Nb$_{9}$O$_{48}$F$_{12}$) with a supercell lattice constant of $\sim$ 10 \AA. We rationalize the choice of such a supercell size because $\sim$ 10 \AA\ is a good cut-off to maintain enough distance between each atom and its periodic images and to encapsulate all cluster-interaction distances. Subsequently, sGCMC simulations are performed for every structure in the ensemble, and for a given chemical potential, the species (e.g., Li) content is computed using the following average
\begin{equation}
    \langle x_{\text{Li}}^{\mu} \rangle = \frac{1}{N_{\text{ens}}} (\Bar{x}_1 + \Bar{x}_2 + ...),
\end{equation}
where $\{\Bar{x}_1, \Bar{x}_2 ...\}$ are the averaged Li content given by the thermally equilibrated sGCMC sample starting from each structure. In this work, we used 30 disordered structures for the ensemble average.

\section{Simulation results}
\begin{figure}[t]
\centering
\includegraphics[width=\linewidth]{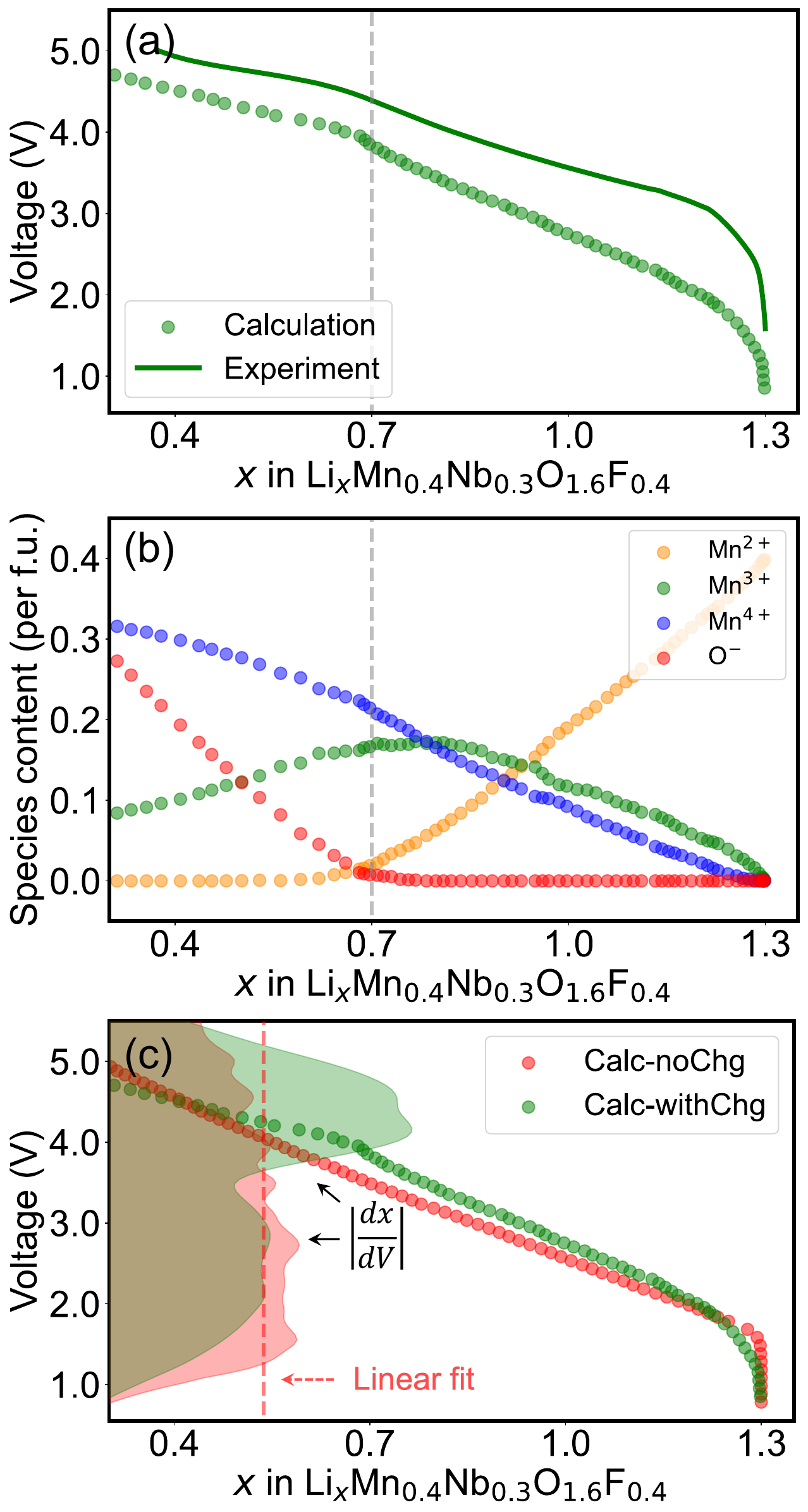}
\caption{(a) Calculated (circle) and experimental (solid line) voltage profiles of (Li$_{1.3-x}$Mn$_{0.4}$Nb$_{0.3}$O$_{1.6}$F$_{0.4}$). (b) Calculated content of Mn$^{2+}$, Mn$^{3+}$, Mn$^{4+}$ and O$^{-}$ per f.u. as a function of Li content ($x$). (c) Comparison of voltage profiles generated with a charge-decorated CE (green dots) and a CE without charge decoration (red dots). The shaded area represents the distribution of the derivative quantity $|\frac{dx}{dV}|$, which reflects the Li-site energy distribution during the intercalation process.}
\label{fig:profile} 
\end{figure}

We demonstrate the procedures presented above to model the intercalation thermodynamics of Li$_{1.3-x}$Mn$_{0.4}$Nb$_{0.3}$O$_{1.6}$F$_{0.4}$ (LMNOF). Figure \ref{fig:profile}(a) presents the simulated and the experimental voltage profiles \cite{cationVacancy_liliang}. The solid green line is the experimental charging profile under a low current density (20 mA/g) between 1.5 and 5.0 V. The bottom of the discharge state (1.5 V) is aligned to be the fully lithiated state (Li$_{1.3}$Mn$_{0.4}$Nb$_{0.3}$O$_{1.6}$F$_{0.4}$).  The green dots represent the computed voltage profile at T = 300 K. The slope and turning point in the slope are in good agreement with the experimental profile for $0.4\leq x\leq1.3$, indicating that the Li/vacancy interaction and redox potentials of Mn and O are well described in our model. In the highly charged region ($x<0.7$) specifically, the simulation shows remarkably good agreement with the experiment reproducing the fact that the slope of the profile becomes flatter at $x\sim0.7$ (marked by the gray dashed line in Fig. \ref{fig:profile}(a)). The computed voltage profile is systematically lower than the experimental one, which is well-known for most DFT functionals that are not augmented with a Hubbard $U$ correction \cite{Zhou2004_LDAU}. Even though the SCAN functional removes more self-interaction than previous LDA and GGA \cite{Sun2015SCAN}, it does not fully remove it and underestimates the intercalation voltage.

The fraction of each redox-active species during intercalation is presented in Fig. \ref{fig:profile}(b). Mn$^{2+/3+/4+}$ are represented by orange/green/blue dots, and O$^-$ are represented by red dots. Comparing Fig. \ref{fig:profile}(a) and (b) makes it apparent that the Li content where the voltage profile flattens ($x\sim0.7$) corresponds to the start of the oxidation of O$^{2-}$ to O$^{-}$. The lowering of the voltage slope as the system changes from TM redox to O redox is consistent with the higher dilution of the oxygen charge compared to TM redox centers. Figure \ref{fig:profile}(b) also reveals several key points about the redox mechanism in LMNOF. The oxidation of Mn does not appear consecutively from  Mn$^{2+}$ to Mn$^{3+}$ to Mn$^{4+}$. The amount of Mn$^{3+}$ and Mn$^{4+}$ increase simultaneously as Mn$^{2+}$ begins to be oxidized. The fact that different oxidation states of Mn co-exist over a wide range of Li content is likely due to the variety of local chemical environments induced by cation disorder. The co-existence is consistent with the marginal stability of Mn$^{3+}$ and its propensity to disproportionate into Mn$^{2}$ and Mn$^{4+}$ when it cannot exist in an environment where it can lower its energy significantly through a Jahn--Teller distortion \cite{Reed2004_review}. O-redox occurs after all the Mn$^{2+}$ has been consumed but before all the Mn atoms are fully oxidized to +4 valence. The hybridized redox mechanism between Mn and O has been confirmed by previous synchrotron characterization experiments \cite{cationVacancy_liliang} and on related materials Li$_{1.2}$Mn$_{0.4}$Ti$_{0.4}$O$_{2.0}$ by mass spectroscopy \cite{Huang2023_Mn_O_redox}.

\section{Discussion}

Typical modeling of the intercalation energetics only includes the Li/vacancy degree of freedom, assuming that the electronic degrees of freedom can be integrated out. Integrating out degrees of freedom is based on two key assumptions \cite{Ceder1993}: (1) One assumes that the degree of freedom that is variationally removed is always optimized in the DFT calculation. More specifically, this would require that for a given Li/vacancy configuration, the DFT calculation can easily find the charge decoration with the lowest energy. While this is a reasonable assumption for systems with a highly delocalized charge such as Li$_x$CoO$_2$ \cite{Menetrier1999_MIT_LiCoO2}, it is unlikely to be the case for disordered systems where the variation of local environments and the highly localized charge easily lead to charge metastability in DFT. (2) The second assumption when degrees of freedom are variationally removed is that their entropy contribution can be neglected, as their variational optimization is supposed to find their ground state. The role of electronic entropy in intercalation systems is yet to be fully understood, but at least in Li$_x$FePO$_4$ it has been shown to be critical to reproduce the correct intercalation behavior \cite{Zhou2006electronic_entropy}. Both assumptions are unlikely to be valid in DRX cathodes, given their multitude of possible redox couples and large variations of energetic environments.

To illustrate the significance of accurately capturing the electronic degree of freedom in DRX materials, we fitted another CE parameterizing only the Li/vacancy interactions. Mn and O atoms are considered single charge-less species regardless of their oxidation states in such a CE. The computed voltage profiles with the charge-decorated CE (in green) and the undecorated CE (in red) are compared in Fig. \ref{fig:profile}(c). The undecorated CE yields a featureless voltage profile and virtually no change of slope as a function of Li content. The mechanism-related details are poorly portrayed in the profile. The shaded areas in Fig. \ref{fig:profile}(c) are the relative derivative of capacity with respect to voltage ($dx/dV$). Two distinct peaks are observed in the charge-decorated calculation (green), indicating the contribution from the Mn-redox and O-redox, whereas they cannot be adequately distinguished in the undecorated version (red).

\begin{figure}[tb]
\centering
\includegraphics[width=\linewidth]{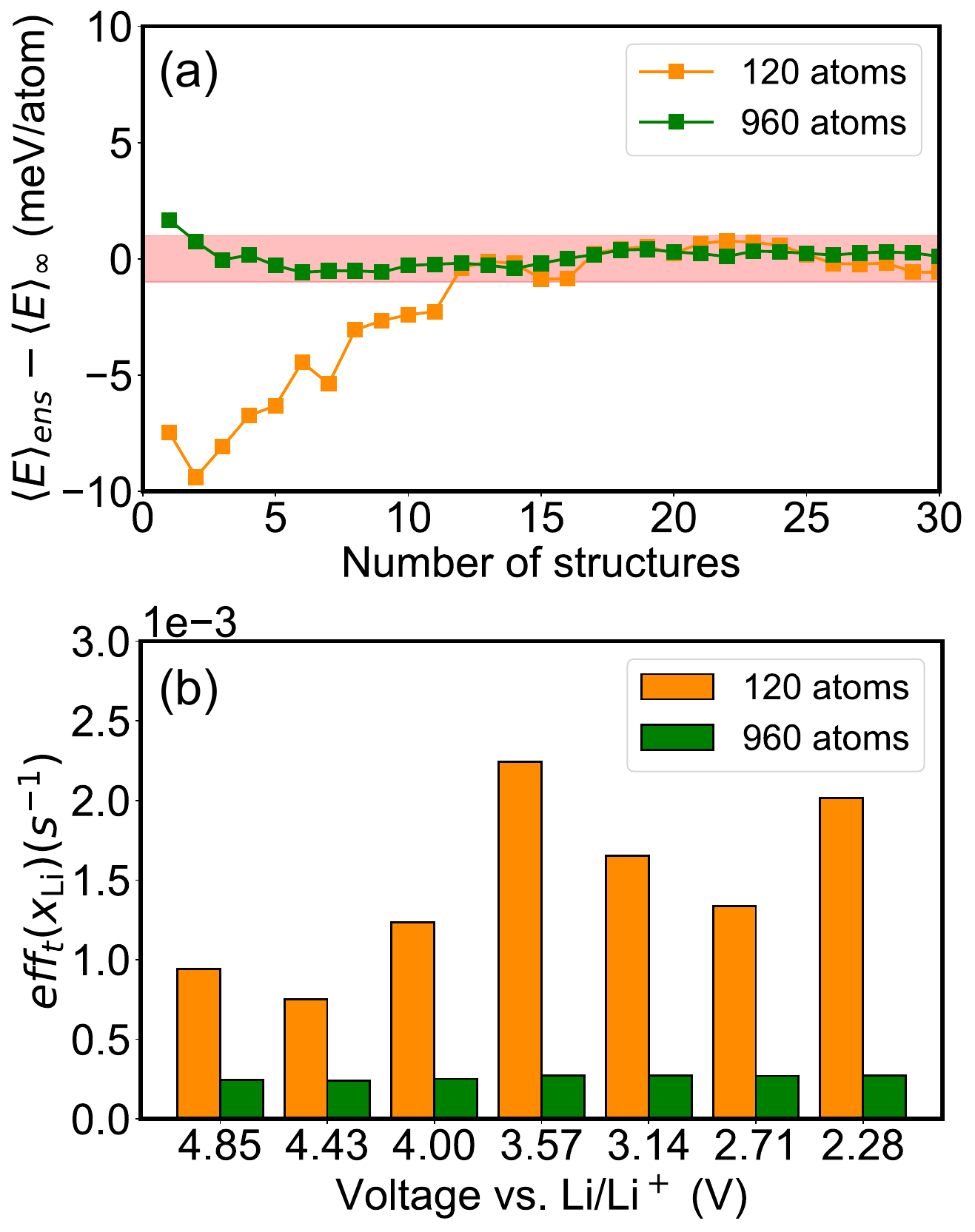}
\caption{(a) Average energy (per atom) among pristine structures as a function of the number of pristine structures selected in an ensemble. The red-shaded region indicates a variation of $\pm 1$ meV/atom of $\langle E\rangle_{\infty}$.  The orange/green lines represent the results of the supercell structure with 120/960 atoms, respectively. (b) Comparison of computational efficiency (defined in Eq. \eqref{eq:teff}) between supercells with 120 and 960 atoms, computed and averaged over all the pristine structures in the ensembles. }
\label{fig:convergence_efficiency}
\end{figure}

\begin{figure*}[tb]
\centering
\includegraphics[width=\linewidth]{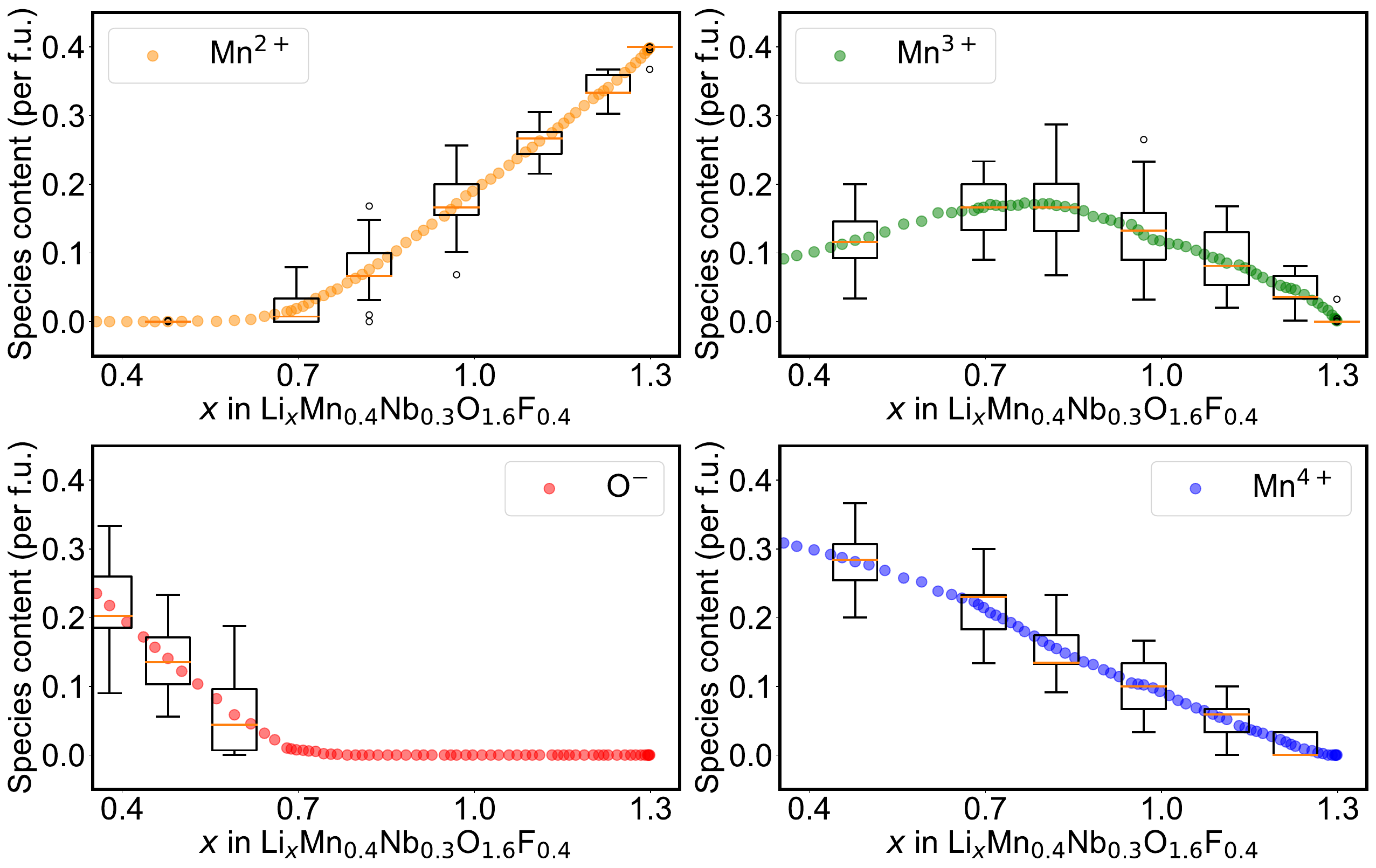}
\caption{Contents of each redox species per formula unit as a function of the Li content $x$ in Li$_{x}$Mn$_{0.4}$Nb$_{0.3}$O$_{1.6}$F$_{0.4}$ as computed from sGCMC simulations at T = 300 K. The statistics are computed based on an ensemble of disordered structures of in supercells of Li$_{30x}$Mn$_{12}$Nb$_{9}$O$_{48}$F$_{12}$. The average values over different disordered structures are marked with colored dots, the median values are marked with orange lines, and the variances are marked with error bars around the medians.}
\label{fig:SRO-var-species} 
\end{figure*}

We also highlight the importance of using the ensemble average method to capture the multitude of environments in disordered systems. In principle, one can approach the representation of disordered systems by using a single structure in a very large supercell (e.g., special quasi-random structure (SQS) approach \cite{Zunger1990_SQS}) or by making an ensemble of multiple smaller-sized supercell structures and taking an average of the computed quantities. These two approaches are statistically equivalent, given that enough structures have been used with the smaller-sized supercells.  Figure \ref{fig:convergence_efficiency}(a) shows the number of the structures required for the average energy per atom over structures to convergence within $\pm 1$ meV/atom of the equilibrium average $\langle E \rangle_\infty$ when choosing from an ensemble of representative structures in canonical MC \cite{cationVacancy_liliang}. The green line represents the results in a supercell of 960 atoms, whereas the orange line represents the results in a supercell of 120 atoms. The smaller-sized supercell requires drawing $\sim$ ten times more structures to converge, with the number of structures required roughly scaling with the supercell size ($960 = 120 \times 8)$. A large supercell approaches the actual distribution of SRO with fewer structures. However, a much longer MC sampling time is required in a large supercell, whereas in a smaller supercell, one can easily benefit from parallelizing multiple pristine structures to require much less total time consumption. To illustrate this factor, we introduce computational efficiency as a quantitative benchmark for sGCMC. Figure \ref{fig:convergence_efficiency}(b) shows the computational efficiency in sGCMC at varied voltages in supercells with 120 (orange) and 960 (green) atoms. Higher efficiency indicates that less computational time is required (see details about the definition in the Appendix). Approximating the computational efficiency, the sampling time required in each pristine structure scales $\sim O(N^1)$ with the supercell size. Therefore, with the help of parallelization, using an ensemble of smaller structures is statistically as accurate but practically more efficient than using fewer large structures. An ensemble of smaller structures is also more tractable in DFT for computing other properties when necessary, such as electronic structure, whereas DFT is computationally prohibitive in large supercells even when the SRO can be well presented. 

Finally, we like to discuss the necessity of using enough structure samples when smaller-sized supercells are used to model disordered systems. Figure \ref{fig:SRO-var-species} shows the variance of redox-active species with respect to the averaged Li content.  As a result of the disorder, the variance in the content of each species among pristine structures is not negligible. This further discourages the methods based on convex-hull and direct DFT evaluations (e.g., GS-algo) to accurately determine voltage profiles for materials with significant disorder such as DRX. When the size of the supercell in which one enumerates possible Li/vacancy and charge configurations is too small or the number of structure samples is too few, as is often the case in GS-algo, the variance resulting from different local chemical environments cannot be adequately captured.

\section{Summary \& Outlook}
In DRX materials, the abundance of available redox-active transition metals creates new opportunities for cathode design, such as using the redox reactions between $\text{Cr}^{3+}/ \text{Cr}^{4+}/\text{Cr}^{6+}$ \cite{Huang2021_Cr, Lun2020_high_entropy}, $\text{V}^{3+} / \text{V}^{4+}/\text{V}^{5+}$ \cite{Nakajima2017_V_DRX}, or even $\text{Fe}^{2+}/ \text{Fe}^{3+}/\text{Fe}^{4+}$ \cite{lebens2021electrochemical}. However, the high number of chemical components, distinguished by their valence states in DRX materials renders the “curse of dimensionality” (CoD) the main simulation obstacle \cite{weinan2020machine}. As the complexity of the energy model grows with the number of components and charge decoration, the computational cost grows exponentially fast. For example, CoD has been an essential impediment for the computational design of high-entropy cathodes \cite{zhang2022high_entropy_NMC, Lun2020_high_entropy}. 

In this work, we propose the following procedure to obtain accurate intercalation voltage profiles in DRX with multi-redox reactions: (1) build a training set containing different fully lithiated structures and enumerate the Li/vacancy orderings at varied delithiation levels calculated 
 by DFT; 
(2) construct a charge-decorated CE Hamiltonian and fit the ECIs using the sparse regression technique to address the fast growth of the cluster basis and rank-deficiency issues (e.g., $\ell_0\ell_2$-norm regularized regression); and (3) run sGCMC in an ensemble of disordered structures under charge balance to obtain physically rational energetics and compute the voltage profiles.

As demonstrated in this work, the workflow above provides an effective way to study the intercalation chemistry of oxides with Mn and O redox activity. Looking forward, the recent development of machine-learning force-fields (MLFF) may provide new opportunities to accelerate the training structure generation process by accurately approximating DFT calculations \cite{Xie2022_MLFF_CE,chen2022_m3gnet, deng2023chgnet, Takamoto2022_PFP}. The CE Hamiltonian can be further obtained by coarse-graining the MLFF predicted configurational energy, especially using the charge-informed MLFF to include the heterovalent states of transition metals \cite{deng2023chgnet}. We believe such an approach has the potential to bridge up first-principles calculations, force fields, and cluster expansions and give a higher accuracy sampling at a lower computational cost to study intercalation chemistries in energy storage materials.

\section{Acknowledgements}
This work was funded by the U.S. Department of Energy, Office of Science, Office of Basic Energy Sciences, Materials Sciences and Engineering Division under Contract No. DE-AC0205CH11231 (Materials Project program KC23MP). The work was also supported by the computational resources provided by the Extreme Science and Engineering Discovery Environment (XSEDE), supported by National Science Foundation grant number ACI1053575; the National Energy Research Scientific Computing Center (NERSC), a U.S. Department of Energy Office of Science User Facility located at Lawrence Berkeley National Laboratory; and the Lawrencium Computational Cluster resource provided by the IT Division at the Lawrence Berkeley National Laboratory.

\appendix
\numberwithin{equation}{section}
\section{DFT Calculations}\label{appendix:DFT}
DFT calculations were performed with the \textit{Vienna ab initio simulation package} (VASP) using the projector-augmented wave method \cite{kresse1996VASP, kresse1999PAW}, a plane-wave basis set with an energy cutoff of 520 eV, and a reciprocal space discretization of 25 \textit{k}-points per \AA$^{-1}$. All the calculations were converged to $10^{-6}$ eV in total energy for electronic loops and 0.02 eV/Å in interatomic forces for ionic loops. To model the Li-Mn-Nb-O-F oxyfluoride system, we relied on the regularized strongly constrained and appropriately normed meta-GGA exchange-correlation functional (r$^2$SCAN) \cite{Sun2015SCAN,furness2020r2SCAN}, which is believed to better capture the cation--anion hybridization and Li coordination preference \cite{zhang2018_npjSCAN}. r$^2$SCAN provides better computational efficiency than the earlier version of SCAN \cite{kingsbury2022r2SCAN_PRM}.

\section{Table Exchange in sGCMC}\label{appdix:GCMC}

Because the TE Metropolis steps are proposed by randomly selecting and exchanging species in a configuration, the forward and backward proposal probability for a TE step can be imbalanced. To ensure detailed balance in the simulation, the acceptance probability of a TE step is adjusted as follows:
\begin{equation}
\label{eq:importance}
\begin{aligned}
p_{\bm{\sigma \sigma'}} &= \min \left\{1, r \right\},\\
r &= \frac{ \prod_{u_s \neq 0} n_s !}{ \prod_{u_s \neq 0} (n_s + u_s)!}\exp \left[-\frac{\left(\Delta E_{\boldsymbol{\sigma \sigma'}} - \sum_{s} \mu_s u_s\right)}{k_B T}  \right],
\end{aligned}
\end{equation}
where $\boldsymbol{\sigma}$ and $\boldsymbol{\sigma'}$ are the configurations before and after applying the TE step, respectively; $n_s$ is the number of species $s$ in configuration $\bm{\sigma}$ before the step; $u_s$ represents the change of species $s$ amount after the step; and the vector value $\boldsymbol{u} = (u_1, u_2, ..., u_s)$ must fall into one of the three aforementioned TE categories or their inverses. All TE categories and their inverses were selected with equal probability.

Additionally, in the TE sGCMC, a percentage ($w=20\%$) of canonical swaps between the Li/vacancy and different oxidation states of the same elements are mixed with TE steps to improve the efficiency of exploring different Li and electronic configurations within the same compositions. For each chemical potential in sGCMC, we applied simulated annealing from T = 5000 K to 100 K to approach the Li/vacancy and oxidation states ground-state configuration. Then, starting from such a ground state, the sGCMC was run at T = 300 K with 500,000 steps to thermally equilibrate the system, and another 500,000 steps were used to generate a sample of configurations for analysis \cite{xie2022grand}.

\section{Computational Efficiency}\label{appdix:efficiency}

At each Metropolis step $p$ in semigrand-canonical MC, the Li content ($x_{\text{Li}}$) at the current configuration is recorded as $x_{\text{Li}, p}$. We denote $\overline{x_{\text{Li}, [p, q]}}$ as the average of $x_{\text{Li}}$ in a block from simulation step $p$ to step $q$. After thermal equilibration, we define the \emph{block mean variance} at block length $L$, $\text{Var}(\overline{x_{\text{Li}}}_L)$, as the variance of the \emph{block means} $\overline{x_{\text{Li}, [p, p+L]}}, \overline{x_{\text{Li}, [p+L, p+2L]}}, ...$ for each block containing $L$ samples. The block mean variance can be used as a measure of uncertainty when estimating $\overline{\theta}$ using a block mean. With the block mean variance, we further define the \emph{computational efficiency},
\begin{equation}
\label{eq:teff}
    \text{eff}_t(x_{\text{Li}}) = \frac{\tau ^ 2}{\overline{T}_L\text{Var}(\overline{x_{\text{Li}}}_L)},
\end{equation}
where $\tau^2$ is the ensemble variance of Li content $x_{\text{Li}}$ approximated by the variance of all MC steps after thermal equilibration, $\overline{T}_L$ is the average CPU time spent in each $L$-steps block. Given the same set of hardware used in MC simulations, the higher the computational efficiency, the less sampling time required for the uncertainty of the average Li content (i.e., the block mean variance) to be decreased below the same threshold. In brief, higher sampling efficiency means less sampling time.

\bibliography{reference.bib}

\begin{thebibliography}{66}%
\makeatletter
\providecommand \@ifxundefined [1]{%
 \@ifx{#1\undefined}
}%
\providecommand \@ifnum [1]{%
 \ifnum #1\expandafter \@firstoftwo
 \else \expandafter \@secondoftwo
 \fi
}%
\providecommand \@ifx [1]{%
 \ifx #1\expandafter \@firstoftwo
 \else \expandafter \@secondoftwo
 \fi
}%
\providecommand \natexlab [1]{#1}%
\providecommand \enquote  [1]{``#1''}%
\providecommand \bibnamefont  [1]{#1}%
\providecommand \bibfnamefont [1]{#1}%
\providecommand \citenamefont [1]{#1}%
\providecommand \href@noop [0]{\@secondoftwo}%
\providecommand \href [0]{\begingroup \@sanitize@url \@href}%
\providecommand \@href[1]{\@@startlink{#1}\@@href}%
\providecommand \@@href[1]{\endgroup#1\@@endlink}%
\providecommand \@sanitize@url [0]{\catcode `\\12\catcode `\$12\catcode
  `\&12\catcode `\#12\catcode `\^12\catcode `\_12\catcode `\%12\relax}%
\providecommand \@@startlink[1]{}%
\providecommand \@@endlink[0]{}%
\providecommand \url  [0]{\begingroup\@sanitize@url \@url }%
\providecommand \@url [1]{\endgroup\@href {#1}{\urlprefix }}%
\providecommand \urlprefix  [0]{URL }%
\providecommand \Eprint [0]{\href }%
\providecommand \doibase [0]{https://doi.org/}%
\providecommand \selectlanguage [0]{\@gobble}%
\providecommand \bibinfo  [0]{\@secondoftwo}%
\providecommand \bibfield  [0]{\@secondoftwo}%
\providecommand \translation [1]{[#1]}%
\providecommand \BibitemOpen [0]{}%
\providecommand \bibitemStop [0]{}%
\providecommand \bibitemNoStop [0]{.\EOS\space}%
\providecommand \EOS [0]{\spacefactor3000\relax}%
\providecommand \BibitemShut  [1]{\csname bibitem#1\endcsname}%
\let\auto@bib@innerbib\@empty
\bibitem [{\citenamefont {Olivetti}\ \emph {et~al.}(2017)\citenamefont
  {Olivetti}, \citenamefont {Ceder}, \citenamefont {Gaustad},\ and\
  \citenamefont {Fu}}]{Olivetti2017}%
  \BibitemOpen
  \bibfield  {author} {\bibinfo {author} {\bibfnamefont {E.~A.}\ \bibnamefont
  {Olivetti}}, \bibinfo {author} {\bibfnamefont {G.}~\bibnamefont {Ceder}},
  \bibinfo {author} {\bibfnamefont {G.~G.}\ \bibnamefont {Gaustad}},\ and\
  \bibinfo {author} {\bibfnamefont {X.}~\bibnamefont {Fu}},\ }\bibfield
  {title} {\bibinfo {title} {{Lithium-Ion Battery Supply Chain Considerations:
  Analysis of Potential Bottlenecks in Critical Metals}},\ }\href
  {https://doi.org/10.1016/j.joule.2017.08.019} {\bibfield  {journal} {\bibinfo
   {journal} {Joule}\ }\textbf {\bibinfo {volume} {1}},\ \bibinfo {pages} {229}
  (\bibinfo {year} {2017})}\BibitemShut {NoStop}%
\bibitem [{\citenamefont {Goodenough}\ and\ \citenamefont
  {Kim}(2010)}]{Goodenough2010}%
  \BibitemOpen
  \bibfield  {author} {\bibinfo {author} {\bibfnamefont {J.~B.}\ \bibnamefont
  {Goodenough}}\ and\ \bibinfo {author} {\bibfnamefont {Y.}~\bibnamefont
  {Kim}},\ }\bibfield  {title} {\bibinfo {title} {{Challenges for Rechargeable
  Li Batteries}},\ }\href {https://doi.org/10.1021/cm901452z} {\bibfield
  {journal} {\bibinfo  {journal} {Chemistry of Materials}\ }\textbf {\bibinfo
  {volume} {22}},\ \bibinfo {pages} {587} (\bibinfo {year} {2010})}\BibitemShut
  {NoStop}%
\bibitem [{\citenamefont {Ji}\ \emph {et~al.}(2019{\natexlab{a}})\citenamefont
  {Ji}, \citenamefont {Urban}, \citenamefont {Kitchaev}, \citenamefont {Kwon},
  \citenamefont {Artrith}, \citenamefont {Ophus}, \citenamefont {Huang},
  \citenamefont {Cai}, \citenamefont {Shi}, \citenamefont {Kim}, \citenamefont
  {Kim},\ and\ \citenamefont {Ceder}}]{Ji2019_NatComm_SRO}%
  \BibitemOpen
  \bibfield  {author} {\bibinfo {author} {\bibfnamefont {H.}~\bibnamefont
  {Ji}}, \bibinfo {author} {\bibfnamefont {A.}~\bibnamefont {Urban}}, \bibinfo
  {author} {\bibfnamefont {D.~A.}\ \bibnamefont {Kitchaev}}, \bibinfo {author}
  {\bibfnamefont {D.-H.}\ \bibnamefont {Kwon}}, \bibinfo {author}
  {\bibfnamefont {N.}~\bibnamefont {Artrith}}, \bibinfo {author} {\bibfnamefont
  {C.}~\bibnamefont {Ophus}}, \bibinfo {author} {\bibfnamefont
  {W.}~\bibnamefont {Huang}}, \bibinfo {author} {\bibfnamefont
  {Z.}~\bibnamefont {Cai}}, \bibinfo {author} {\bibfnamefont {T.}~\bibnamefont
  {Shi}}, \bibinfo {author} {\bibfnamefont {J.~C.}\ \bibnamefont {Kim}},
  \bibinfo {author} {\bibfnamefont {H.}~\bibnamefont {Kim}},\ and\ \bibinfo
  {author} {\bibfnamefont {G.}~\bibnamefont {Ceder}},\ }\bibfield  {title}
  {\bibinfo {title} {{Hidden structural and chemical order controls lithium
  transport in cation-disordered oxides for rechargeable batteries}},\ }\href
  {https://doi.org/10.1038/s41467-019-08490-w} {\bibfield  {journal} {\bibinfo
  {journal} {Nature Communications}\ }\textbf {\bibinfo {volume} {10}},\
  \bibinfo {pages} {592} (\bibinfo {year} {2019}{\natexlab{a}})}\BibitemShut
  {NoStop}%
\bibitem [{\citenamefont {Richards}\ \emph {et~al.}(2018)\citenamefont
  {Richards}, \citenamefont {Dacek}, \citenamefont {Kitchaev},\ and\
  \citenamefont {Ceder}}]{Richards2018_fluorination}%
  \BibitemOpen
  \bibfield  {author} {\bibinfo {author} {\bibfnamefont {W.~D.}\ \bibnamefont
  {Richards}}, \bibinfo {author} {\bibfnamefont {S.~T.}\ \bibnamefont {Dacek}},
  \bibinfo {author} {\bibfnamefont {D.~A.}\ \bibnamefont {Kitchaev}},\ and\
  \bibinfo {author} {\bibfnamefont {G.}~\bibnamefont {Ceder}},\ }\bibfield
  {title} {\bibinfo {title} {{Fluorination of Lithium‐Excess Transition Metal
  Oxide Cathode Materials}},\ }\href {https://doi.org/10.1002/aenm.201701533}
  {\bibfield  {journal} {\bibinfo  {journal} {Advanced Energy Materials}\
  }\textbf {\bibinfo {volume} {8}},\ \bibinfo {pages} {1701533} (\bibinfo
  {year} {2018})}\BibitemShut {NoStop}%
\bibitem [{\citenamefont {Huang}\ \emph {et~al.}(2022)\citenamefont {Huang},
  \citenamefont {Zhong}, \citenamefont {Ha}, \citenamefont {Lun}, \citenamefont
  {Tian}, \citenamefont {Balasubramanian}, \citenamefont {Yang},\ and\
  \citenamefont {Ceder}}]{Huang2022_oxyvac}%
  \BibitemOpen
  \bibfield  {author} {\bibinfo {author} {\bibfnamefont {J.}~\bibnamefont
  {Huang}}, \bibinfo {author} {\bibfnamefont {P.}~\bibnamefont {Zhong}},
  \bibinfo {author} {\bibfnamefont {Y.}~\bibnamefont {Ha}}, \bibinfo {author}
  {\bibfnamefont {Z.}~\bibnamefont {Lun}}, \bibinfo {author} {\bibfnamefont
  {Y.}~\bibnamefont {Tian}}, \bibinfo {author} {\bibfnamefont {M.}~\bibnamefont
  {Balasubramanian}}, \bibinfo {author} {\bibfnamefont {W.}~\bibnamefont
  {Yang}},\ and\ \bibinfo {author} {\bibfnamefont {G.}~\bibnamefont {Ceder}},\
  }\bibfield  {title} {\bibinfo {title} {{Oxygen Vacancy Introduction to
  Increase the Capacity and Voltage Retention in Li‐Excess Cathode
  Materials}},\ }\href {https://doi.org/10.1002/sstr.202200343} {\bibfield
  {journal} {\bibinfo  {journal} {Small Structures}\ ,\ \bibinfo {pages}
  {2200343}} (\bibinfo {year} {2022})}\BibitemShut {NoStop}%
\bibitem [{\citenamefont {Zhang}\ \emph {et~al.}(2022)\citenamefont {Zhang},
  \citenamefont {Wang}, \citenamefont {Zou}, \citenamefont {Lin}, \citenamefont
  {Ma}, \citenamefont {Yin}, \citenamefont {Li}, \citenamefont {Xu},
  \citenamefont {Jia}, \citenamefont {Li}, \citenamefont {Sainio},
  \citenamefont {Kisslinger}, \citenamefont {Trask}, \citenamefont {Ehrlich},
  \citenamefont {Yang}, \citenamefont {Kiss}, \citenamefont {Ge}, \citenamefont
  {Polzin}, \citenamefont {Lee}, \citenamefont {Xu}, \citenamefont {Ren},\ and\
  \citenamefont {Xin}}]{zhang2022high_entropy_NMC}%
  \BibitemOpen
  \bibfield  {author} {\bibinfo {author} {\bibfnamefont {R.}~\bibnamefont
  {Zhang}}, \bibinfo {author} {\bibfnamefont {C.}~\bibnamefont {Wang}},
  \bibinfo {author} {\bibfnamefont {P.}~\bibnamefont {Zou}}, \bibinfo {author}
  {\bibfnamefont {R.}~\bibnamefont {Lin}}, \bibinfo {author} {\bibfnamefont
  {L.}~\bibnamefont {Ma}}, \bibinfo {author} {\bibfnamefont {L.}~\bibnamefont
  {Yin}}, \bibinfo {author} {\bibfnamefont {T.}~\bibnamefont {Li}}, \bibinfo
  {author} {\bibfnamefont {W.}~\bibnamefont {Xu}}, \bibinfo {author}
  {\bibfnamefont {H.}~\bibnamefont {Jia}}, \bibinfo {author} {\bibfnamefont
  {Q.}~\bibnamefont {Li}}, \bibinfo {author} {\bibfnamefont {S.}~\bibnamefont
  {Sainio}}, \bibinfo {author} {\bibfnamefont {K.}~\bibnamefont {Kisslinger}},
  \bibinfo {author} {\bibfnamefont {S.~E.}\ \bibnamefont {Trask}}, \bibinfo
  {author} {\bibfnamefont {S.~N.}\ \bibnamefont {Ehrlich}}, \bibinfo {author}
  {\bibfnamefont {Y.}~\bibnamefont {Yang}}, \bibinfo {author} {\bibfnamefont
  {A.~M.}\ \bibnamefont {Kiss}}, \bibinfo {author} {\bibfnamefont
  {M.}~\bibnamefont {Ge}}, \bibinfo {author} {\bibfnamefont {B.~J.}\
  \bibnamefont {Polzin}}, \bibinfo {author} {\bibfnamefont {S.~J.}\
  \bibnamefont {Lee}}, \bibinfo {author} {\bibfnamefont {W.}~\bibnamefont
  {Xu}}, \bibinfo {author} {\bibfnamefont {Y.}~\bibnamefont {Ren}},\ and\
  \bibinfo {author} {\bibfnamefont {H.~L.}\ \bibnamefont {Xin}},\ }\bibfield
  {title} {\bibinfo {title} {{Compositionally complex doping for zero-strain
  zero-cobalt layered cathodes}},\ }\href
  {https://doi.org/10.1038/s41586-022-05115-z} {\bibfield  {journal} {\bibinfo
  {journal} {Nature}\ }\textbf {\bibinfo {volume} {610}},\ \bibinfo {pages}
  {67} (\bibinfo {year} {2022})}\BibitemShut {NoStop}%
\bibitem [{\citenamefont {Zhou}\ \emph {et~al.}(2022)\citenamefont {Zhou},
  \citenamefont {Li}, \citenamefont {Ha}, \citenamefont {Zhang}, \citenamefont
  {Dachraoui}, \citenamefont {Liu}, \citenamefont {Zhang}, \citenamefont {Liu},
  \citenamefont {Liu}, \citenamefont {Battaglia}, \citenamefont {Yang},
  \citenamefont {Liu},\ and\ \citenamefont {Yang}}]{Zhou2022_zero_strain}%
  \BibitemOpen
  \bibfield  {author} {\bibinfo {author} {\bibfnamefont {K.}~\bibnamefont
  {Zhou}}, \bibinfo {author} {\bibfnamefont {Y.}~\bibnamefont {Li}}, \bibinfo
  {author} {\bibfnamefont {Y.}~\bibnamefont {Ha}}, \bibinfo {author}
  {\bibfnamefont {M.}~\bibnamefont {Zhang}}, \bibinfo {author} {\bibfnamefont
  {W.}~\bibnamefont {Dachraoui}}, \bibinfo {author} {\bibfnamefont
  {H.}~\bibnamefont {Liu}}, \bibinfo {author} {\bibfnamefont {C.}~\bibnamefont
  {Zhang}}, \bibinfo {author} {\bibfnamefont {X.}~\bibnamefont {Liu}}, \bibinfo
  {author} {\bibfnamefont {F.}~\bibnamefont {Liu}}, \bibinfo {author}
  {\bibfnamefont {C.}~\bibnamefont {Battaglia}}, \bibinfo {author}
  {\bibfnamefont {W.}~\bibnamefont {Yang}}, \bibinfo {author} {\bibfnamefont
  {J.}~\bibnamefont {Liu}},\ and\ \bibinfo {author} {\bibfnamefont
  {Y.}~\bibnamefont {Yang}},\ }\bibfield  {title} {\bibinfo {title} {{A Nearly
  Zero-Strain Li-Rich Rock-Salt Oxide with Multielectron Redox Reactions as a
  Cathode for Li-Ion Batteries}},\ }\href
  {https://doi.org/10.1021/acs.chemmater.2c02519} {\bibfield  {journal}
  {\bibinfo  {journal} {Chemistry of Materials}\ }\textbf {\bibinfo {volume}
  {34}},\ \bibinfo {pages} {9711} (\bibinfo {year} {2022})}\BibitemShut
  {NoStop}%
\bibitem [{\citenamefont {Zhao}\ \emph {et~al.}(2022)\citenamefont {Zhao},
  \citenamefont {Tian}, \citenamefont {Lun}, \citenamefont {Cai}, \citenamefont
  {Chen}, \citenamefont {Ouyang},\ and\ \citenamefont
  {Ceder}}]{Zhao2022_zerostrain}%
  \BibitemOpen
  \bibfield  {author} {\bibinfo {author} {\bibfnamefont {X.}~\bibnamefont
  {Zhao}}, \bibinfo {author} {\bibfnamefont {Y.}~\bibnamefont {Tian}}, \bibinfo
  {author} {\bibfnamefont {Z.}~\bibnamefont {Lun}}, \bibinfo {author}
  {\bibfnamefont {Z.}~\bibnamefont {Cai}}, \bibinfo {author} {\bibfnamefont
  {T.}~\bibnamefont {Chen}}, \bibinfo {author} {\bibfnamefont {B.}~\bibnamefont
  {Ouyang}},\ and\ \bibinfo {author} {\bibfnamefont {G.}~\bibnamefont
  {Ceder}},\ }\bibfield  {title} {\bibinfo {title} {{Design principles for
  zero-strain Li-ion cathodes}},\ }\href
  {https://doi.org/10.1016/j.joule.2022.05.018} {\bibfield  {journal} {\bibinfo
   {journal} {Joule}\ }\textbf {\bibinfo {volume} {6}},\ \bibinfo {pages}
  {1654} (\bibinfo {year} {2022})}\BibitemShut {NoStop}%
\bibitem [{\citenamefont {Meng}\ and\ \citenamefont {{Arroyo-De
  Dompablo}}(2009)}]{Meng2009_review}%
  \BibitemOpen
  \bibfield  {author} {\bibinfo {author} {\bibfnamefont {Y.~S.}\ \bibnamefont
  {Meng}}\ and\ \bibinfo {author} {\bibfnamefont {M.~E.}\ \bibnamefont
  {{Arroyo-De Dompablo}}},\ }\bibfield  {title} {\bibinfo {title} {{First
  principles computational materials design for energy storage materials in
  lithium ion batteries}},\ }\href {https://doi.org/10.1039/b901825e}
  {\bibfield  {journal} {\bibinfo  {journal} {Energy and Environmental
  Science}\ }\textbf {\bibinfo {volume} {2}},\ \bibinfo {pages} {589} (\bibinfo
  {year} {2009})}\BibitemShut {NoStop}%
\bibitem [{\citenamefont {Urban}\ \emph {et~al.}(2016)\citenamefont {Urban},
  \citenamefont {Seo},\ and\ \citenamefont {Ceder}}]{Urban2016npj}%
  \BibitemOpen
  \bibfield  {author} {\bibinfo {author} {\bibfnamefont {A.}~\bibnamefont
  {Urban}}, \bibinfo {author} {\bibfnamefont {D.-H.}\ \bibnamefont {Seo}},\
  and\ \bibinfo {author} {\bibfnamefont {G.}~\bibnamefont {Ceder}},\ }\bibfield
   {title} {\bibinfo {title} {{Computational understanding of Li-ion
  batteries}},\ }\href {https://doi.org/10.1038/npjcompumats.2016.2} {\bibfield
   {journal} {\bibinfo  {journal} {npj Computational Materials}\ }\textbf
  {\bibinfo {volume} {2}},\ \bibinfo {pages} {16002} (\bibinfo {year}
  {2016})}\BibitemShut {NoStop}%
\bibitem [{\citenamefont {{Van Der Ven}}\ \emph {et~al.}(2020)\citenamefont
  {{Van Der Ven}}, \citenamefont {Deng}, \citenamefont {Banerjee},\ and\
  \citenamefont {Ong}}]{VanDerVen2020}%
  \BibitemOpen
  \bibfield  {author} {\bibinfo {author} {\bibfnamefont {A.}~\bibnamefont {{Van
  Der Ven}}}, \bibinfo {author} {\bibfnamefont {Z.}~\bibnamefont {Deng}},
  \bibinfo {author} {\bibfnamefont {S.}~\bibnamefont {Banerjee}},\ and\
  \bibinfo {author} {\bibfnamefont {S.~P.}\ \bibnamefont {Ong}},\ }\bibfield
  {title} {\bibinfo {title} {{Rechargeable Alkali-Ion Battery Materials: Theory
  and Computation}},\ }\href {https://doi.org/10.1021/acs.chemrev.9b00601}
  {\bibfield  {journal} {\bibinfo  {journal} {Chemical Reviews}\ }\textbf
  {\bibinfo {volume} {120}},\ \bibinfo {pages} {6977} (\bibinfo {year}
  {2020})}\BibitemShut {NoStop}%
\bibitem [{\citenamefont {Aydinol}\ \emph {et~al.}(1997)\citenamefont
  {Aydinol}, \citenamefont {Kohan}, \citenamefont {Ceder}, \citenamefont
  {Cho},\ and\ \citenamefont {Joannopoulos}}]{aydinol1997_abinit_voltage}%
  \BibitemOpen
  \bibfield  {author} {\bibinfo {author} {\bibfnamefont {M.~K.}\ \bibnamefont
  {Aydinol}}, \bibinfo {author} {\bibfnamefont {A.~F.}\ \bibnamefont {Kohan}},
  \bibinfo {author} {\bibfnamefont {G.}~\bibnamefont {Ceder}}, \bibinfo
  {author} {\bibfnamefont {K.}~\bibnamefont {Cho}},\ and\ \bibinfo {author}
  {\bibfnamefont {J.}~\bibnamefont {Joannopoulos}},\ }\bibfield  {title}
  {\bibinfo {title} {{Ab initio study of lithium intercalation in metal oxides
  and metal dichalcogenides}},\ }\href
  {https://doi.org/10.1103/PhysRevB.56.1354} {\bibfield  {journal} {\bibinfo
  {journal} {Physical Review B}\ }\textbf {\bibinfo {volume} {56}},\ \bibinfo
  {pages} {1354} (\bibinfo {year} {1997})}\BibitemShut {NoStop}%
\bibitem [{\citenamefont {Urban}\ \emph {et~al.}(2017)\citenamefont {Urban},
  \citenamefont {Abdellahi}, \citenamefont {Dacek}, \citenamefont {Artrith},\
  and\ \citenamefont {Ceder}}]{Urban2017_PRL}%
  \BibitemOpen
  \bibfield  {author} {\bibinfo {author} {\bibfnamefont {A.}~\bibnamefont
  {Urban}}, \bibinfo {author} {\bibfnamefont {A.}~\bibnamefont {Abdellahi}},
  \bibinfo {author} {\bibfnamefont {S.}~\bibnamefont {Dacek}}, \bibinfo
  {author} {\bibfnamefont {N.}~\bibnamefont {Artrith}},\ and\ \bibinfo {author}
  {\bibfnamefont {G.}~\bibnamefont {Ceder}},\ }\bibfield  {title} {\bibinfo
  {title} {{Electronic-Structure Origin of Cation Disorder in Transition-Metal
  Oxides}},\ }\href {https://doi.org/10.1103/PhysRevLett.119.176402} {\bibfield
   {journal} {\bibinfo  {journal} {Physical Review Letters}\ }\textbf {\bibinfo
  {volume} {119}},\ \bibinfo {pages} {1} (\bibinfo {year} {2017})}\BibitemShut
  {NoStop}%
\bibitem [{\citenamefont {Lun}\ \emph {et~al.}(2019)\citenamefont {Lun},
  \citenamefont {Ouyang}, \citenamefont {Kitchaev}, \citenamefont
  {Cl{\'{e}}ment}, \citenamefont {Papp}, \citenamefont {Balasubramanian},
  \citenamefont {Tian}, \citenamefont {Lei}, \citenamefont {Shi}, \citenamefont
  {McCloskey}, \citenamefont {Lee},\ and\ \citenamefont
  {Ceder}}]{Lun2019AEM_F_cycle}%
  \BibitemOpen
  \bibfield  {author} {\bibinfo {author} {\bibfnamefont {Z.}~\bibnamefont
  {Lun}}, \bibinfo {author} {\bibfnamefont {B.}~\bibnamefont {Ouyang}},
  \bibinfo {author} {\bibfnamefont {D.~A.}\ \bibnamefont {Kitchaev}}, \bibinfo
  {author} {\bibfnamefont {R.~J.}\ \bibnamefont {Cl{\'{e}}ment}}, \bibinfo
  {author} {\bibfnamefont {J.~K.}\ \bibnamefont {Papp}}, \bibinfo {author}
  {\bibfnamefont {M.}~\bibnamefont {Balasubramanian}}, \bibinfo {author}
  {\bibfnamefont {Y.}~\bibnamefont {Tian}}, \bibinfo {author} {\bibfnamefont
  {T.}~\bibnamefont {Lei}}, \bibinfo {author} {\bibfnamefont {T.}~\bibnamefont
  {Shi}}, \bibinfo {author} {\bibfnamefont {B.~D.}\ \bibnamefont {McCloskey}},
  \bibinfo {author} {\bibfnamefont {J.}~\bibnamefont {Lee}},\ and\ \bibinfo
  {author} {\bibfnamefont {G.}~\bibnamefont {Ceder}},\ }\bibfield  {title}
  {\bibinfo {title} {{Improved Cycling Performance of Li-Excess
  Cation-Disordered Cathode Materials upon Fluorine Substitution}},\ }\href
  {https://doi.org/10.1002/aenm.201802959} {\bibfield  {journal} {\bibinfo
  {journal} {Advanced Energy Materials}\ }\textbf {\bibinfo {volume} {9}},\
  \bibinfo {pages} {1802959} (\bibinfo {year} {2019})}\BibitemShut {NoStop}%
\bibitem [{\citenamefont {Zhou}\ \emph {et~al.}(2006)\citenamefont {Zhou},
  \citenamefont {Maxisch},\ and\ \citenamefont
  {Ceder}}]{Zhou2006electronic_entropy}%
  \BibitemOpen
  \bibfield  {author} {\bibinfo {author} {\bibfnamefont {F.}~\bibnamefont
  {Zhou}}, \bibinfo {author} {\bibfnamefont {T.}~\bibnamefont {Maxisch}},\ and\
  \bibinfo {author} {\bibfnamefont {G.}~\bibnamefont {Ceder}},\ }\bibfield
  {title} {\bibinfo {title} {{Configurational Electronic Entropy and the Phase
  Diagram of Mixed-Valence Oxides: The Case of Li$_x$FePO$_4$}},\ }\href
  {https://doi.org/10.1103/PhysRevLett.97.155704} {\bibfield  {journal}
  {\bibinfo  {journal} {Physical Review Letters}\ }\textbf {\bibinfo {volume}
  {97}},\ \bibinfo {pages} {155704} (\bibinfo {year} {2006})}\BibitemShut
  {NoStop}%
\bibitem [{\citenamefont {Lee}\ \emph {et~al.}(2021)\citenamefont {Lee},
  \citenamefont {Wang}, \citenamefont {Malik}, \citenamefont {Dong},
  \citenamefont {Huang}, \citenamefont {Seo},\ and\ \citenamefont
  {Li}}]{Lee2021_critivity}%
  \BibitemOpen
  \bibfield  {author} {\bibinfo {author} {\bibfnamefont {J.}~\bibnamefont
  {Lee}}, \bibinfo {author} {\bibfnamefont {C.}~\bibnamefont {Wang}}, \bibinfo
  {author} {\bibfnamefont {R.}~\bibnamefont {Malik}}, \bibinfo {author}
  {\bibfnamefont {Y.}~\bibnamefont {Dong}}, \bibinfo {author} {\bibfnamefont
  {Y.}~\bibnamefont {Huang}}, \bibinfo {author} {\bibfnamefont
  {D.}~\bibnamefont {Seo}},\ and\ \bibinfo {author} {\bibfnamefont
  {J.}~\bibnamefont {Li}},\ }\bibfield  {title} {\bibinfo {title} {{Determining
  the Criticality of Li‐Excess for Disordered‐Rocksalt Li‐Ion Battery
  Cathodes}},\ }\href {https://doi.org/10.1002/aenm.202100204} {\bibfield
  {journal} {\bibinfo  {journal} {Advanced Energy Materials}\ }\textbf
  {\bibinfo {volume} {11}},\ \bibinfo {pages} {2100204} (\bibinfo {year}
  {2021})}\BibitemShut {NoStop}%
\bibitem [{\citenamefont {Li}\ \emph {et~al.}(2021)\citenamefont {Li},
  \citenamefont {Sougrati}, \citenamefont {Rousse}, \citenamefont {Morozov},
  \citenamefont {Dedryv{\`{e}}re}, \citenamefont {Iadecola}, \citenamefont
  {Senyshyn}, \citenamefont {Zhang}, \citenamefont {Abakumov}, \citenamefont
  {Doublet},\ and\ \citenamefont {Tarascon}}]{Li2021_NatureChem}%
  \BibitemOpen
  \bibfield  {author} {\bibinfo {author} {\bibfnamefont {B.}~\bibnamefont
  {Li}}, \bibinfo {author} {\bibfnamefont {M.~T.}\ \bibnamefont {Sougrati}},
  \bibinfo {author} {\bibfnamefont {G.}~\bibnamefont {Rousse}}, \bibinfo
  {author} {\bibfnamefont {A.~V.}\ \bibnamefont {Morozov}}, \bibinfo {author}
  {\bibfnamefont {R.}~\bibnamefont {Dedryv{\`{e}}re}}, \bibinfo {author}
  {\bibfnamefont {A.}~\bibnamefont {Iadecola}}, \bibinfo {author}
  {\bibfnamefont {A.}~\bibnamefont {Senyshyn}}, \bibinfo {author}
  {\bibfnamefont {L.}~\bibnamefont {Zhang}}, \bibinfo {author} {\bibfnamefont
  {A.~M.}\ \bibnamefont {Abakumov}}, \bibinfo {author} {\bibfnamefont {M.-L.}\
  \bibnamefont {Doublet}},\ and\ \bibinfo {author} {\bibfnamefont {J.-M.}\
  \bibnamefont {Tarascon}},\ }\bibfield  {title} {\bibinfo {title}
  {{Correlating ligand-to-metal charge transfer with voltage hysteresis in a
  Li-rich rock-salt compound exhibiting anionic redox}},\ }\href
  {https://doi.org/10.1038/s41557-021-00775-2} {\bibfield  {journal} {\bibinfo
  {journal} {Nature Chemistry}\ }\textbf {\bibinfo {volume} {13}},\ \bibinfo
  {pages} {1070} (\bibinfo {year} {2021})}\BibitemShut {NoStop}%
\bibitem [{\citenamefont {Yao}\ \emph {et~al.}(2018)\citenamefont {Yao},
  \citenamefont {Kim}, \citenamefont {He}, \citenamefont {Hegde},\ and\
  \citenamefont {Wolverton}}]{Yao2018_LiMnO}%
  \BibitemOpen
  \bibfield  {author} {\bibinfo {author} {\bibfnamefont {Z.}~\bibnamefont
  {Yao}}, \bibinfo {author} {\bibfnamefont {S.}~\bibnamefont {Kim}}, \bibinfo
  {author} {\bibfnamefont {J.}~\bibnamefont {He}}, \bibinfo {author}
  {\bibfnamefont {V.~I.}\ \bibnamefont {Hegde}},\ and\ \bibinfo {author}
  {\bibfnamefont {C.}~\bibnamefont {Wolverton}},\ }\bibfield  {title} {\bibinfo
  {title} {{Interplay of cation and anion redox in Li$_4$Mn$_2$O$_5$ cathode
  material and prediction of improved Li$_4$(Mn,M)$_2$O$_5$ electrodes for
  Li-ion batteries}},\ }\href {https://doi.org/10.1126/sciadv.aao6754}
  {\bibfield  {journal} {\bibinfo  {journal} {Science Advances}\ }\textbf
  {\bibinfo {volume} {4}},\ \bibinfo {pages} {eaao6754} (\bibinfo {year}
  {2018})}\BibitemShut {NoStop}%
\bibitem [{\citenamefont {Li}\ \emph {et~al.}(2018)\citenamefont {Li},
  \citenamefont {Wu}, \citenamefont {Yao}, \citenamefont {Xu}, \citenamefont
  {Thackeray}, \citenamefont {Wolverton},\ and\ \citenamefont
  {Dravid}}]{li2018LiCo_NonEqui}%
  \BibitemOpen
  \bibfield  {author} {\bibinfo {author} {\bibfnamefont {Q.}~\bibnamefont
  {Li}}, \bibinfo {author} {\bibfnamefont {J.}~\bibnamefont {Wu}}, \bibinfo
  {author} {\bibfnamefont {Z.}~\bibnamefont {Yao}}, \bibinfo {author}
  {\bibfnamefont {Y.}~\bibnamefont {Xu}}, \bibinfo {author} {\bibfnamefont
  {M.~M.}\ \bibnamefont {Thackeray}}, \bibinfo {author} {\bibfnamefont
  {C.}~\bibnamefont {Wolverton}},\ and\ \bibinfo {author} {\bibfnamefont
  {V.~P.}\ \bibnamefont {Dravid}},\ }\bibfield  {title} {\bibinfo {title}
  {{Dynamic imaging of metastable reaction pathways in lithiated cobalt oxide
  electrodes}},\ }\href {https://doi.org/10.1016/j.nanoen.2017.11.052}
  {\bibfield  {journal} {\bibinfo  {journal} {Nano Energy}\ }\textbf {\bibinfo
  {volume} {44}},\ \bibinfo {pages} {15} (\bibinfo {year} {2018})}\BibitemShut
  {NoStop}%
\bibitem [{\citenamefont {Ji}\ \emph {et~al.}(2019{\natexlab{b}})\citenamefont
  {Ji}, \citenamefont {Kitchaev}, \citenamefont {Lun}, \citenamefont {Kim},
  \citenamefont {Foley}, \citenamefont {Kwon}, \citenamefont {Tian},
  \citenamefont {Balasubramanian}, \citenamefont {Bianchini}, \citenamefont
  {Cai}, \citenamefont {Cl{\'{e}}ment}, \citenamefont {Kim},\ and\
  \citenamefont {Ceder}}]{Ji2019_Ni}%
  \BibitemOpen
  \bibfield  {author} {\bibinfo {author} {\bibfnamefont {H.}~\bibnamefont
  {Ji}}, \bibinfo {author} {\bibfnamefont {D.~A.}\ \bibnamefont {Kitchaev}},
  \bibinfo {author} {\bibfnamefont {Z.}~\bibnamefont {Lun}}, \bibinfo {author}
  {\bibfnamefont {H.}~\bibnamefont {Kim}}, \bibinfo {author} {\bibfnamefont
  {E.}~\bibnamefont {Foley}}, \bibinfo {author} {\bibfnamefont {D.~H.}\
  \bibnamefont {Kwon}}, \bibinfo {author} {\bibfnamefont {Y.}~\bibnamefont
  {Tian}}, \bibinfo {author} {\bibfnamefont {M.}~\bibnamefont
  {Balasubramanian}}, \bibinfo {author} {\bibfnamefont {M.}~\bibnamefont
  {Bianchini}}, \bibinfo {author} {\bibfnamefont {Z.}~\bibnamefont {Cai}},
  \bibinfo {author} {\bibfnamefont {R.~J.}\ \bibnamefont {Cl{\'{e}}ment}},
  \bibinfo {author} {\bibfnamefont {J.~C.}\ \bibnamefont {Kim}},\ and\ \bibinfo
  {author} {\bibfnamefont {G.}~\bibnamefont {Ceder}},\ }\bibfield  {title}
  {\bibinfo {title} {{Computational Investigation and Experimental Realization
  of Disordered High-Capacity Li-Ion Cathodes Based on Ni Redox}},\ }\href
  {https://doi.org/10.1021/acs.chemmater.8b05096} {\bibfield  {journal}
  {\bibinfo  {journal} {Chemistry of Materials}\ }\textbf {\bibinfo {volume}
  {31}},\ \bibinfo {pages} {2431} (\bibinfo {year}
  {2019}{\natexlab{b}})}\BibitemShut {NoStop}%
\bibitem [{\citenamefont {McColl}\ \emph {et~al.}(2022)\citenamefont {McColl},
  \citenamefont {House}, \citenamefont {Rees}, \citenamefont {Squires},
  \citenamefont {Coles}, \citenamefont {Bruce}, \citenamefont {Morgan},\ and\
  \citenamefont {Islam}}]{McColl2022_NatComm}%
  \BibitemOpen
  \bibfield  {author} {\bibinfo {author} {\bibfnamefont {K.}~\bibnamefont
  {McColl}}, \bibinfo {author} {\bibfnamefont {R.~A.}\ \bibnamefont {House}},
  \bibinfo {author} {\bibfnamefont {G.~J.}\ \bibnamefont {Rees}}, \bibinfo
  {author} {\bibfnamefont {A.~G.}\ \bibnamefont {Squires}}, \bibinfo {author}
  {\bibfnamefont {S.~W.}\ \bibnamefont {Coles}}, \bibinfo {author}
  {\bibfnamefont {P.~G.}\ \bibnamefont {Bruce}}, \bibinfo {author}
  {\bibfnamefont {B.~J.}\ \bibnamefont {Morgan}},\ and\ \bibinfo {author}
  {\bibfnamefont {M.~S.}\ \bibnamefont {Islam}},\ }\bibfield  {title} {\bibinfo
  {title} {{Transition metal migration and O$_2$ formation underpin voltage
  hysteresis in oxygen-redox disordered rocksalt cathodes}},\ }\href
  {https://doi.org/10.1038/s41467-022-32983-w} {\bibfield  {journal} {\bibinfo
  {journal} {Nature Communications}\ }\textbf {\bibinfo {volume} {13}},\
  \bibinfo {pages} {5275} (\bibinfo {year} {2022})}\BibitemShut {NoStop}%
\bibitem [{\citenamefont {{\AA}ngqvist}\ \emph {et~al.}(2019)\citenamefont
  {{\AA}ngqvist}, \citenamefont {Mu{\~n}oz}, \citenamefont {Rahm},
  \citenamefont {Fransson}, \citenamefont {Durniak}, \citenamefont {Rozyczko},
  \citenamefont {Rod},\ and\ \citenamefont {Erhart}}]{aangqvist2019icet}%
  \BibitemOpen
  \bibfield  {author} {\bibinfo {author} {\bibfnamefont {M.}~\bibnamefont
  {{\AA}ngqvist}}, \bibinfo {author} {\bibfnamefont {W.~A.}\ \bibnamefont
  {Mu{\~n}oz}}, \bibinfo {author} {\bibfnamefont {J.~M.}\ \bibnamefont {Rahm}},
  \bibinfo {author} {\bibfnamefont {E.}~\bibnamefont {Fransson}}, \bibinfo
  {author} {\bibfnamefont {C.}~\bibnamefont {Durniak}}, \bibinfo {author}
  {\bibfnamefont {P.}~\bibnamefont {Rozyczko}}, \bibinfo {author}
  {\bibfnamefont {T.~H.}\ \bibnamefont {Rod}},\ and\ \bibinfo {author}
  {\bibfnamefont {P.}~\bibnamefont {Erhart}},\ }\bibfield  {title} {\bibinfo
  {title} {{Icet--a python library for constructing and sampling alloy cluster
  expansions}},\ }\href
  {https://onlinelibrary.wiley.com/doi/10.1002/adts.201900015} {\bibfield
  {journal} {\bibinfo  {journal} {Advanced Theory and Simulations}\ }\textbf
  {\bibinfo {volume} {2}},\ \bibinfo {pages} {1900015} (\bibinfo {year}
  {2019})}\BibitemShut {NoStop}%
\bibitem [{\citenamefont {Wu}\ \emph {et~al.}(2016)\citenamefont {Wu},
  \citenamefont {He}, \citenamefont {Song}, \citenamefont {Gao},\ and\
  \citenamefont {Shi}}]{ShiSiQi2016_CE_review}%
  \BibitemOpen
  \bibfield  {author} {\bibinfo {author} {\bibfnamefont {Q.}~\bibnamefont
  {Wu}}, \bibinfo {author} {\bibfnamefont {B.}~\bibnamefont {He}}, \bibinfo
  {author} {\bibfnamefont {T.}~\bibnamefont {Song}}, \bibinfo {author}
  {\bibfnamefont {J.}~\bibnamefont {Gao}},\ and\ \bibinfo {author}
  {\bibfnamefont {S.}~\bibnamefont {Shi}},\ }\bibfield  {title} {\bibinfo
  {title} {Cluster expansion method and its application in computational
  materials science},\ }\href {https://doi.org/10.1016/j.commatsci.2016.08.034}
  {\bibfield  {journal} {\bibinfo  {journal} {Computational Materials Science}\
  }\textbf {\bibinfo {volume} {125}},\ \bibinfo {pages} {243} (\bibinfo {year}
  {2016})}\BibitemShut {NoStop}%
\bibitem [{\citenamefont {van~de Walle}\ \emph {et~al.}(2002)\citenamefont
  {van~de Walle}, \citenamefont {Asta},\ and\ \citenamefont
  {Ceder}}]{van2002_ATAT}%
  \BibitemOpen
  \bibfield  {author} {\bibinfo {author} {\bibfnamefont {A.}~\bibnamefont
  {van~de Walle}}, \bibinfo {author} {\bibfnamefont {M.}~\bibnamefont {Asta}},\
  and\ \bibinfo {author} {\bibfnamefont {G.}~\bibnamefont {Ceder}},\ }\bibfield
   {title} {\bibinfo {title} {{The alloy theoretic automated toolkit: A user
  guide}},\ }\href
  {https://www.sciencedirect.com/science/article/pii/S0364591602800062}
  {\bibfield  {journal} {\bibinfo  {journal} {Calphad}\ }\textbf {\bibinfo
  {volume} {26}},\ \bibinfo {pages} {539} (\bibinfo {year} {2002})}\BibitemShut
  {NoStop}%
\bibitem [{\citenamefont {Seko}\ and\ \citenamefont
  {Tanaka}(2014)}]{Seko2014_longrange}%
  \BibitemOpen
  \bibfield  {author} {\bibinfo {author} {\bibfnamefont {A.}~\bibnamefont
  {Seko}}\ and\ \bibinfo {author} {\bibfnamefont {I.}~\bibnamefont {Tanaka}},\
  }\bibfield  {title} {\bibinfo {title} {{Cluster expansion of multicomponent
  ionic systems with controlled accuracy: importance of long-range interactions
  in heterovalent ionic systems}},\ }\href
  {https://doi.org/10.1088/0953-8984/26/11/115403} {\bibfield  {journal}
  {\bibinfo  {journal} {Journal of Physics: Condensed Matter}\ }\textbf
  {\bibinfo {volume} {26}},\ \bibinfo {pages} {115403} (\bibinfo {year}
  {2014})}\BibitemShut {NoStop}%
\bibitem [{\citenamefont {Barroso-Luque}\ \emph
  {et~al.}(2022{\natexlab{a}})\citenamefont {Barroso-Luque}, \citenamefont
  {Zhong}, \citenamefont {Yang}, \citenamefont {Xie}, \citenamefont {Chen},
  \citenamefont {Ouyang},\ and\ \citenamefont
  {Ceder}}]{Barroso-Luque2022_theory}%
  \BibitemOpen
  \bibfield  {author} {\bibinfo {author} {\bibfnamefont {L.}~\bibnamefont
  {Barroso-Luque}}, \bibinfo {author} {\bibfnamefont {P.}~\bibnamefont
  {Zhong}}, \bibinfo {author} {\bibfnamefont {J.~H.}\ \bibnamefont {Yang}},
  \bibinfo {author} {\bibfnamefont {F.}~\bibnamefont {Xie}}, \bibinfo {author}
  {\bibfnamefont {T.}~\bibnamefont {Chen}}, \bibinfo {author} {\bibfnamefont
  {B.}~\bibnamefont {Ouyang}},\ and\ \bibinfo {author} {\bibfnamefont
  {G.}~\bibnamefont {Ceder}},\ }\bibfield  {title} {\bibinfo {title} {{Cluster
  expansions of multicomponent ionic materials: Formalism and methodology}},\
  }\href {https://doi.org/10.1103/PhysRevB.106.144202} {\bibfield  {journal}
  {\bibinfo  {journal} {Physical Review B}\ }\textbf {\bibinfo {volume}
  {106}},\ \bibinfo {pages} {144202} (\bibinfo {year}
  {2022}{\natexlab{a}})}\BibitemShut {NoStop}%
\bibitem [{\citenamefont {van~de Walle}(2009)}]{VandeWalle2009}%
  \BibitemOpen
  \bibfield  {author} {\bibinfo {author} {\bibfnamefont {A.}~\bibnamefont
  {van~de Walle}},\ }\bibfield  {title} {\bibinfo {title} {{Multicomponent
  multisublattice alloys, nonconfigurational entropy and other additions to the
  Alloy Theoretic Automated Toolkit}},\ }\href
  {https://doi.org/10.1016/j.calphad.2008.12.005} {\bibfield  {journal}
  {\bibinfo  {journal} {Calphad}\ }\textbf {\bibinfo {volume} {33}},\ \bibinfo
  {pages} {266} (\bibinfo {year} {2009})}\BibitemShut {NoStop}%
\bibitem [{\citenamefont {Nelson}\ \emph {et~al.}(2013)\citenamefont {Nelson},
  \citenamefont {Hart}, \citenamefont {Zhou},\ and\ \citenamefont
  {Ozoliņ{\v{s}}}}]{Nelson2013}%
  \BibitemOpen
  \bibfield  {author} {\bibinfo {author} {\bibfnamefont {L.~J.}\ \bibnamefont
  {Nelson}}, \bibinfo {author} {\bibfnamefont {G.~L.~W.}\ \bibnamefont {Hart}},
  \bibinfo {author} {\bibfnamefont {F.}~\bibnamefont {Zhou}},\ and\ \bibinfo
  {author} {\bibfnamefont {V.}~\bibnamefont {Ozoliņ{\v{s}}}},\ }\bibfield
  {title} {\bibinfo {title} {{Compressive sensing as a paradigm for building
  physics models}},\ }\href {https://doi.org/10.1103/PhysRevB.87.035125}
  {\bibfield  {journal} {\bibinfo  {journal} {Physical Review B}\ }\textbf
  {\bibinfo {volume} {87}},\ \bibinfo {pages} {035125} (\bibinfo {year}
  {2013})}\BibitemShut {NoStop}%
\bibitem [{\citenamefont {Seko}\ \emph {et~al.}(2014)\citenamefont {Seko},
  \citenamefont {Takahashi},\ and\ \citenamefont {Tanaka}}]{Seko_sparseL1}%
  \BibitemOpen
  \bibfield  {author} {\bibinfo {author} {\bibfnamefont {A.}~\bibnamefont
  {Seko}}, \bibinfo {author} {\bibfnamefont {A.}~\bibnamefont {Takahashi}},\
  and\ \bibinfo {author} {\bibfnamefont {I.}~\bibnamefont {Tanaka}},\
  }\bibfield  {title} {\bibinfo {title} {Sparse representation for a potential
  energy surface},\ }\href {https://doi.org/10.1103/PhysRevB.90.024101}
  {\bibfield  {journal} {\bibinfo  {journal} {Phys. Rev. B}\ }\textbf {\bibinfo
  {volume} {90}},\ \bibinfo {pages} {024101} (\bibinfo {year}
  {2014})}\BibitemShut {NoStop}%
\bibitem [{\citenamefont {Zhong}\ \emph {et~al.}(2022)\citenamefont {Zhong},
  \citenamefont {Chen}, \citenamefont {Barroso-Luque}, \citenamefont {Xie},\
  and\ \citenamefont {Ceder}}]{Zhong2022L0L2}%
  \BibitemOpen
  \bibfield  {author} {\bibinfo {author} {\bibfnamefont {P.}~\bibnamefont
  {Zhong}}, \bibinfo {author} {\bibfnamefont {T.}~\bibnamefont {Chen}},
  \bibinfo {author} {\bibfnamefont {L.}~\bibnamefont {Barroso-Luque}}, \bibinfo
  {author} {\bibfnamefont {F.}~\bibnamefont {Xie}},\ and\ \bibinfo {author}
  {\bibfnamefont {G.}~\bibnamefont {Ceder}},\ }\bibfield  {title} {\bibinfo
  {title} {{An $\ell_0\ell_2$-norm regularized regression model for
  construction of robust cluster expansions in multicomponent systems}},\
  }\href {https://doi.org/10.1103/PhysRevB.106.024203} {\bibfield  {journal}
  {\bibinfo  {journal} {Physical Review B}\ }\textbf {\bibinfo {volume}
  {106}},\ \bibinfo {pages} {024203} (\bibinfo {year} {2022})}\BibitemShut
  {NoStop}%
\bibitem [{\citenamefont {Puchala}\ \emph {et~al.}(2023)\citenamefont
  {Puchala}, \citenamefont {Thomas}, \citenamefont {Natarajan}, \citenamefont
  {Goiri}, \citenamefont {Behara}, \citenamefont {Kaufman},\ and\ \citenamefont
  {{Van der Ven}}}]{Puchala2023_CASM}%
  \BibitemOpen
  \bibfield  {author} {\bibinfo {author} {\bibfnamefont {B.}~\bibnamefont
  {Puchala}}, \bibinfo {author} {\bibfnamefont {J.~C.}\ \bibnamefont {Thomas}},
  \bibinfo {author} {\bibfnamefont {A.~R.}\ \bibnamefont {Natarajan}}, \bibinfo
  {author} {\bibfnamefont {J.~G.}\ \bibnamefont {Goiri}}, \bibinfo {author}
  {\bibfnamefont {S.~S.}\ \bibnamefont {Behara}}, \bibinfo {author}
  {\bibfnamefont {J.~L.}\ \bibnamefont {Kaufman}},\ and\ \bibinfo {author}
  {\bibfnamefont {A.}~\bibnamefont {{Van der Ven}}},\ }\bibfield  {title}
  {\bibinfo {title} {{CASM — A software package for first-principles based
  study of multicomponent crystalline solids}},\ }\href
  {https://doi.org/10.1016/j.commatsci.2022.111897} {\bibfield  {journal}
  {\bibinfo  {journal} {Computational Materials Science}\ }\textbf {\bibinfo
  {volume} {217}},\ \bibinfo {pages} {111897} (\bibinfo {year}
  {2023})}\BibitemShut {NoStop}%
\bibitem [{\citenamefont {Van~der Ven}\ \emph {et~al.}(1998)\citenamefont
  {Van~der Ven}, \citenamefont {Aydinol}, \citenamefont {Ceder}, \citenamefont
  {Kresse},\ and\ \citenamefont {Hafner}}]{VanderVen1998_LiCoO2}%
  \BibitemOpen
  \bibfield  {author} {\bibinfo {author} {\bibfnamefont {A.}~\bibnamefont
  {Van~der Ven}}, \bibinfo {author} {\bibfnamefont {M.~K.}\ \bibnamefont
  {Aydinol}}, \bibinfo {author} {\bibfnamefont {G.}~\bibnamefont {Ceder}},
  \bibinfo {author} {\bibfnamefont {G.}~\bibnamefont {Kresse}},\ and\ \bibinfo
  {author} {\bibfnamefont {J.}~\bibnamefont {Hafner}},\ }\bibfield  {title}
  {\bibinfo {title} {{First-principles investigation of phase stability in
  Li$_x$CoO$_2$}},\ }\href {http://link.aps.org/doi/10.1103/PhysRevB.58.2975}
  {\bibfield  {journal} {\bibinfo  {journal} {Physical Review B}\ }\textbf
  {\bibinfo {volume} {58}},\ \bibinfo {pages} {2975} (\bibinfo {year}
  {1998})}\BibitemShut {NoStop}%
\bibitem [{\citenamefont {Wolverton}\ and\ \citenamefont
  {Zunger}(1998)}]{Wolverton1998_LiCO2}%
  \BibitemOpen
  \bibfield  {author} {\bibinfo {author} {\bibfnamefont {C.}~\bibnamefont
  {Wolverton}}\ and\ \bibinfo {author} {\bibfnamefont {A.}~\bibnamefont
  {Zunger}},\ }\bibfield  {title} {\bibinfo {title} {{First-principles
  prediction of vacancy order-disorder and intercalation battery voltages in
  Li$_x$CoO$_2$}},\ }\href {https://doi.org/10.1103/PhysRevLett.81.606}
  {\bibfield  {journal} {\bibinfo  {journal} {Physical Review Letters}\
  }\textbf {\bibinfo {volume} {81}},\ \bibinfo {pages} {606} (\bibinfo {year}
  {1998})}\BibitemShut {NoStop}%
\bibitem [{\citenamefont {Houchins}\ and\ \citenamefont
  {Viswanathan}(2020)}]{houchins2020_MLP_gcmc}%
  \BibitemOpen
  \bibfield  {author} {\bibinfo {author} {\bibfnamefont {G.}~\bibnamefont
  {Houchins}}\ and\ \bibinfo {author} {\bibfnamefont {V.}~\bibnamefont
  {Viswanathan}},\ }\bibfield  {title} {\bibinfo {title} {{An accurate
  machine-learning calculator for optimization of Li-ion battery cathodes}},\
  }\href {https://aip.scitation.org/doi/10.1063/5.0015872} {\bibfield
  {journal} {\bibinfo  {journal} {The Journal of Chemical Physics}\ }\textbf
  {\bibinfo {volume} {153}},\ \bibinfo {pages} {054124} (\bibinfo {year}
  {2020})}\BibitemShut {NoStop}%
\bibitem [{\citenamefont {Kolli}\ and\ \citenamefont {{Van Der
  Ven}}(2018)}]{Kolli2018_MgTiS2}%
  \BibitemOpen
  \bibfield  {author} {\bibinfo {author} {\bibfnamefont {S.~K.}\ \bibnamefont
  {Kolli}}\ and\ \bibinfo {author} {\bibfnamefont {A.}~\bibnamefont {{Van Der
  Ven}}},\ }\bibfield  {title} {\bibinfo {title} {{First-Principles Study of
  Spinel MgTiS$_2$ as a Cathode Material}},\ }\href
  {https://doi.org/10.1021/acs.chemmater.8b00552} {\bibfield  {journal}
  {\bibinfo  {journal} {Chemistry of Materials}\ }\textbf {\bibinfo {volume}
  {30}},\ \bibinfo {pages} {2436} (\bibinfo {year} {2018})}\BibitemShut
  {NoStop}%
\bibitem [{\citenamefont {Guo}\ \emph {et~al.}(2023)\citenamefont {Guo},
  \citenamefont {Chen},\ and\ \citenamefont {Ong}}]{guo2022intercalation}%
  \BibitemOpen
  \bibfield  {author} {\bibinfo {author} {\bibfnamefont {X.}~\bibnamefont
  {Guo}}, \bibinfo {author} {\bibfnamefont {C.}~\bibnamefont {Chen}},\ and\
  \bibinfo {author} {\bibfnamefont {S.~P.}\ \bibnamefont {Ong}},\ }\bibfield
  {title} {\bibinfo {title} {{Intercalation Chemistry of the Disordered
  Rocksalt Li$_3$V$_2$O$_5$ Anode from Cluster Expansions and Machine Learning
  Interatomic Potentials}},\ }\href
  {https://doi.org/10.1021/acs.chemmater.2c02839} {\bibfield  {journal}
  {\bibinfo  {journal} {Chemistry of Materials}\ }\textbf {\bibinfo {volume}
  {35}},\ \bibinfo {pages} {1537} (\bibinfo {year} {2023})}\BibitemShut
  {NoStop}%
\bibitem [{\citenamefont {Valleau}\ and\ \citenamefont
  {Cohen}(1980)}]{valleau1980_chg_GCMC}%
  \BibitemOpen
  \bibfield  {author} {\bibinfo {author} {\bibfnamefont {J.~P.}\ \bibnamefont
  {Valleau}}\ and\ \bibinfo {author} {\bibfnamefont {L.~K.}\ \bibnamefont
  {Cohen}},\ }\bibfield  {title} {\bibinfo {title} {{Primitive model
  electrolytes. I. Grand canonical Monte Carlo computations}},\ }\href
  {https://doi.org/10.1063/1.439092} {\bibfield  {journal} {\bibinfo  {journal}
  {The Journal of Chemical Physics}\ }\textbf {\bibinfo {volume} {72}},\
  \bibinfo {pages} {5935} (\bibinfo {year} {1980})}\BibitemShut {NoStop}%
\bibitem [{\citenamefont {Deng}\ \emph {et~al.}(2020)\citenamefont {Deng},
  \citenamefont {{Sai Gautam}}, \citenamefont {Kolli}, \citenamefont {Chotard},
  \citenamefont {Cheetham}, \citenamefont {Masquelier},\ and\ \citenamefont
  {Canepa}}]{Deng2020_Na_electrolyte}%
  \BibitemOpen
  \bibfield  {author} {\bibinfo {author} {\bibfnamefont {Z.}~\bibnamefont
  {Deng}}, \bibinfo {author} {\bibfnamefont {G.}~\bibnamefont {{Sai Gautam}}},
  \bibinfo {author} {\bibfnamefont {S.~K.}\ \bibnamefont {Kolli}}, \bibinfo
  {author} {\bibfnamefont {J.~N.}\ \bibnamefont {Chotard}}, \bibinfo {author}
  {\bibfnamefont {A.~K.}\ \bibnamefont {Cheetham}}, \bibinfo {author}
  {\bibfnamefont {C.}~\bibnamefont {Masquelier}},\ and\ \bibinfo {author}
  {\bibfnamefont {P.}~\bibnamefont {Canepa}},\ }\bibfield  {title} {\bibinfo
  {title} {{Phase Behavior in Rhombohedral NaSiCON Electrolytes and
  Electrodes}},\ }\href
  {https://doi.org/10.1021/ACS.CHEMMATER.0C02695/ASSET/IMAGES/LARGE/CM0C02695_0009.JPEG}
  {\bibfield  {journal} {\bibinfo  {journal} {Chemistry of Materials}\ }\textbf
  {\bibinfo {volume} {32}},\ \bibinfo {pages} {7908} (\bibinfo {year}
  {2020})}\BibitemShut {NoStop}%
\bibitem [{\citenamefont {Xie}\ \emph {et~al.}(2023)\citenamefont {Xie},
  \citenamefont {Zhong}, \citenamefont {Barroso-Luque}, \citenamefont
  {Ouyang},\ and\ \citenamefont {Ceder}}]{xie2022grand}%
  \BibitemOpen
  \bibfield  {author} {\bibinfo {author} {\bibfnamefont {F.}~\bibnamefont
  {Xie}}, \bibinfo {author} {\bibfnamefont {P.}~\bibnamefont {Zhong}}, \bibinfo
  {author} {\bibfnamefont {L.}~\bibnamefont {Barroso-Luque}}, \bibinfo {author}
  {\bibfnamefont {B.}~\bibnamefont {Ouyang}},\ and\ \bibinfo {author}
  {\bibfnamefont {G.}~\bibnamefont {Ceder}},\ }\bibfield  {title} {\bibinfo
  {title} {{Semigrand-canonical Monte-Carlo simulation methods for
  charge-decorated cluster expansions}},\ }\href
  {https://doi.org/10.1016/j.commatsci.2022.112000} {\bibfield  {journal}
  {\bibinfo  {journal} {{Computational Materials Science}}\ }\textbf {\bibinfo
  {volume} {218}},\ \bibinfo {pages} {112000} (\bibinfo {year}
  {2023})}\BibitemShut {NoStop}%
\bibitem [{\citenamefont {Yang}\ \emph {et~al.}(2022)\citenamefont {Yang},
  \citenamefont {Chen}, \citenamefont {Barroso-Luque}, \citenamefont {Jadidi},\
  and\ \citenamefont {Ceder}}]{Yang2022_npj}%
  \BibitemOpen
  \bibfield  {author} {\bibinfo {author} {\bibfnamefont {J.~H.}\ \bibnamefont
  {Yang}}, \bibinfo {author} {\bibfnamefont {T.}~\bibnamefont {Chen}}, \bibinfo
  {author} {\bibfnamefont {L.}~\bibnamefont {Barroso-Luque}}, \bibinfo {author}
  {\bibfnamefont {Z.}~\bibnamefont {Jadidi}},\ and\ \bibinfo {author}
  {\bibfnamefont {G.}~\bibnamefont {Ceder}},\ }\bibfield  {title} {\bibinfo
  {title} {{Approaches for handling high-dimensional cluster expansions of
  ionic systems}},\ }\href {https://doi.org/10.1038/s41524-022-00818-3}
  {\bibfield  {journal} {\bibinfo  {journal} {npj Computational Materials}\
  }\textbf {\bibinfo {volume} {8}},\ \bibinfo {pages} {133} (\bibinfo {year}
  {2022})}\BibitemShut {NoStop}%
\bibitem [{\citenamefont {Seo}\ \emph {et~al.}(2016)\citenamefont {Seo},
  \citenamefont {Lee}, \citenamefont {Urban}, \citenamefont {Malik},
  \citenamefont {Kang},\ and\ \citenamefont {Ceder}}]{Seo2016_NatChem}%
  \BibitemOpen
  \bibfield  {author} {\bibinfo {author} {\bibfnamefont {D.-H.}\ \bibnamefont
  {Seo}}, \bibinfo {author} {\bibfnamefont {J.}~\bibnamefont {Lee}}, \bibinfo
  {author} {\bibfnamefont {A.}~\bibnamefont {Urban}}, \bibinfo {author}
  {\bibfnamefont {R.}~\bibnamefont {Malik}}, \bibinfo {author} {\bibfnamefont
  {S.}~\bibnamefont {Kang}},\ and\ \bibinfo {author} {\bibfnamefont
  {G.}~\bibnamefont {Ceder}},\ }\bibfield  {title} {\bibinfo {title} {{The
  structural and chemical origin of the oxygen redox activity in layered and
  cation-disordered Li-excess cathode materials}},\ }\href
  {https://doi.org/10.1038/nchem.2524} {\bibfield  {journal} {\bibinfo
  {journal} {Nature Chemistry}\ }\textbf {\bibinfo {volume} {8}},\ \bibinfo
  {pages} {692} (\bibinfo {year} {2016})}\BibitemShut {NoStop}%
\bibitem [{\citenamefont {Barroso-Luque}\ \emph
  {et~al.}(2022{\natexlab{b}})\citenamefont {Barroso-Luque}, \citenamefont
  {Yang}, \citenamefont {Xie}, \citenamefont {Chen}, \citenamefont {Kam},
  \citenamefont {Jadidi}, \citenamefont {Zhong},\ and\ \citenamefont
  {Ceder}}]{Barroso-Luque2022smol}%
  \BibitemOpen
  \bibfield  {author} {\bibinfo {author} {\bibfnamefont {L.}~\bibnamefont
  {Barroso-Luque}}, \bibinfo {author} {\bibfnamefont {J.~H.}\ \bibnamefont
  {Yang}}, \bibinfo {author} {\bibfnamefont {F.}~\bibnamefont {Xie}}, \bibinfo
  {author} {\bibfnamefont {T.}~\bibnamefont {Chen}}, \bibinfo {author}
  {\bibfnamefont {R.~L.}\ \bibnamefont {Kam}}, \bibinfo {author} {\bibfnamefont
  {Z.}~\bibnamefont {Jadidi}}, \bibinfo {author} {\bibfnamefont
  {P.}~\bibnamefont {Zhong}},\ and\ \bibinfo {author} {\bibfnamefont
  {G.}~\bibnamefont {Ceder}},\ }\bibfield  {title} {\bibinfo {title} {{smol: A
  Python package for cluster expansions and beyond}},\ }\href
  {https://doi.org/10.21105/joss.04504} {\bibfield  {journal} {\bibinfo
  {journal} {Journal of Open Source Software}\ }\textbf {\bibinfo {volume}
  {7}},\ \bibinfo {pages} {4504} (\bibinfo {year}
  {2022}{\natexlab{b}})}\BibitemShut {NoStop}%
\bibitem [{\citenamefont {Abdellahi}\ \emph {et~al.}(2016)\citenamefont
  {Abdellahi}, \citenamefont {Urban}, \citenamefont {Dacek},\ and\
  \citenamefont {Ceder}}]{Abdellahi2016}%
  \BibitemOpen
  \bibfield  {author} {\bibinfo {author} {\bibfnamefont {A.}~\bibnamefont
  {Abdellahi}}, \bibinfo {author} {\bibfnamefont {A.}~\bibnamefont {Urban}},
  \bibinfo {author} {\bibfnamefont {S.}~\bibnamefont {Dacek}},\ and\ \bibinfo
  {author} {\bibfnamefont {G.}~\bibnamefont {Ceder}},\ }\bibfield  {title}
  {\bibinfo {title} {{The Effect of Cation Disorder on the Average Li
  Intercalation Voltage of Transition-Metal Oxides}},\ }\href
  {https://doi.org/10.1021/acs.chemmater.6b00205} {\bibfield  {journal}
  {\bibinfo  {journal} {Chemistry of Materials}\ }\textbf {\bibinfo {volume}
  {28}},\ \bibinfo {pages} {3659} (\bibinfo {year} {2016})}\BibitemShut
  {NoStop}%
\bibitem [{\citenamefont {Squires}\ and\ \citenamefont
  {Scanlon}(2023)}]{Squires2023_SRO}%
  \BibitemOpen
  \bibfield  {author} {\bibinfo {author} {\bibfnamefont {A.~G.}\ \bibnamefont
  {Squires}}\ and\ \bibinfo {author} {\bibfnamefont {D.~O.}\ \bibnamefont
  {Scanlon}},\ }\bibfield  {title} {\bibinfo {title} {{Understanding the limits
  to short-range order suppression in many-component disordered rock salt
  lithium-ion cathode materials}},\ }\href {https://doi.org/10.1039/D3TA02088F}
  {\bibfield  {journal} {\bibinfo  {journal} {Journal of Materials Chemistry
  A}\ }\textbf {\bibinfo {volume} {11}},\ \bibinfo {pages} {13765} (\bibinfo
  {year} {2023})}\BibitemShut {NoStop}%
\bibitem [{\citenamefont {Huang}\ \emph
  {et~al.}(2023{\natexlab{a}})\citenamefont {Huang}, \citenamefont {Zhong},
  \citenamefont {Ha}, \citenamefont {Cai}, \citenamefont {Byeon}, \citenamefont
  {Huang}, \citenamefont {Sun}, \citenamefont {Xie}, \citenamefont {Hau},
  \citenamefont {Kim}, \citenamefont {Balasubramanian}, \citenamefont
  {McCloskey}, \citenamefont {Yang},\ and\ \citenamefont
  {Ceder}}]{cationVacancy_liliang}%
  \BibitemOpen
  \bibfield  {author} {\bibinfo {author} {\bibfnamefont {L.}~\bibnamefont
  {Huang}}, \bibinfo {author} {\bibfnamefont {P.}~\bibnamefont {Zhong}},
  \bibinfo {author} {\bibfnamefont {Y.}~\bibnamefont {Ha}}, \bibinfo {author}
  {\bibfnamefont {Z.}~\bibnamefont {Cai}}, \bibinfo {author} {\bibfnamefont
  {Y.}~\bibnamefont {Byeon}}, \bibinfo {author} {\bibfnamefont
  {T.}~\bibnamefont {Huang}}, \bibinfo {author} {\bibfnamefont
  {Y.}~\bibnamefont {Sun}}, \bibinfo {author} {\bibfnamefont {F.}~\bibnamefont
  {Xie}}, \bibinfo {author} {\bibfnamefont {H.}~\bibnamefont {Hau}}, \bibinfo
  {author} {\bibfnamefont {H.}~\bibnamefont {Kim}}, \bibinfo {author}
  {\bibfnamefont {M.}~\bibnamefont {Balasubramanian}}, \bibinfo {author}
  {\bibfnamefont {B.~D.}\ \bibnamefont {McCloskey}}, \bibinfo {author}
  {\bibfnamefont {W.}~\bibnamefont {Yang}},\ and\ \bibinfo {author}
  {\bibfnamefont {G.}~\bibnamefont {Ceder}},\ }\bibfield  {title} {\bibinfo
  {title} {{Optimizing Li‐Excess Cation‐Disordered Rocksalt Cathode Design
  Through Partial Li Deficiency}},\ }\href
  {https://doi.org/10.1002/aenm.202202345} {\bibfield  {journal} {\bibinfo
  {journal} {Advanced Energy Materials}\ }\textbf {\bibinfo {volume} {13}},\
  \bibinfo {pages} {2202345} (\bibinfo {year}
  {2023}{\natexlab{a}})}\BibitemShut {NoStop}%
\bibitem [{\citenamefont {Zhou}\ \emph {et~al.}(2004)\citenamefont {Zhou},
  \citenamefont {Cococcioni}, \citenamefont {Marianetti}, \citenamefont
  {Morgan},\ and\ \citenamefont {Ceder}}]{Zhou2004_LDAU}%
  \BibitemOpen
  \bibfield  {author} {\bibinfo {author} {\bibfnamefont {F.}~\bibnamefont
  {Zhou}}, \bibinfo {author} {\bibfnamefont {M.}~\bibnamefont {Cococcioni}},
  \bibinfo {author} {\bibfnamefont {C.~A.}\ \bibnamefont {Marianetti}},
  \bibinfo {author} {\bibfnamefont {D.}~\bibnamefont {Morgan}},\ and\ \bibinfo
  {author} {\bibfnamefont {G.}~\bibnamefont {Ceder}},\ }\bibfield  {title}
  {\bibinfo {title} {{First-principles prediction of redox potentials in
  transition-metal compounds with LDA$+U$}},\ }\href
  {https://doi.org/10.1103/PhysRevB.70.235121} {\bibfield  {journal} {\bibinfo
  {journal} {Physical Review B}\ }\textbf {\bibinfo {volume} {70}},\ \bibinfo
  {pages} {235121} (\bibinfo {year} {2004})},\ \Eprint
  {https://arxiv.org/abs/0406382} {0406382} \BibitemShut {NoStop}%
\bibitem [{\citenamefont {Sun}\ \emph {et~al.}(2015)\citenamefont {Sun},
  \citenamefont {Ruzsinszky},\ and\ \citenamefont {Perdew}}]{Sun2015SCAN}%
  \BibitemOpen
  \bibfield  {author} {\bibinfo {author} {\bibfnamefont {J.}~\bibnamefont
  {Sun}}, \bibinfo {author} {\bibfnamefont {A.}~\bibnamefont {Ruzsinszky}},\
  and\ \bibinfo {author} {\bibfnamefont {J.~P.}\ \bibnamefont {Perdew}},\
  }\bibfield  {title} {\bibinfo {title} {{Strongly Constrained and
  Appropriately Normed Semilocal Density Functional}},\ }\href
  {https://doi.org/10.1103/PhysRevLett.115.036402} {\bibfield  {journal}
  {\bibinfo  {journal} {Physical Review Letters}\ }\textbf {\bibinfo {volume}
  {115}},\ \bibinfo {pages} {036402} (\bibinfo {year} {2015})}\BibitemShut
  {NoStop}%
\bibitem [{\citenamefont {Reed}\ and\ \citenamefont
  {Ceder}(2004)}]{Reed2004_review}%
  \BibitemOpen
  \bibfield  {author} {\bibinfo {author} {\bibfnamefont {J.}~\bibnamefont
  {Reed}}\ and\ \bibinfo {author} {\bibfnamefont {G.}~\bibnamefont {Ceder}},\
  }\bibfield  {title} {\bibinfo {title} {{Role of Electronic Structure in the
  Susceptibility of Metastable Transition-Metal Oxide Structures to
  Transformation}},\ }\href {https://doi.org/10.1021/cr020733x} {\bibfield
  {journal} {\bibinfo  {journal} {Chemical Reviews}\ }\textbf {\bibinfo
  {volume} {104}},\ \bibinfo {pages} {4513} (\bibinfo {year}
  {2004})}\BibitemShut {NoStop}%
\bibitem [{\citenamefont {Huang}\ \emph
  {et~al.}(2023{\natexlab{b}})\citenamefont {Huang}, \citenamefont {Cai},
  \citenamefont {Crafton}, \citenamefont {Kaufman}, \citenamefont {Konz},
  \citenamefont {Bergstrom}, \citenamefont {Kedzie}, \citenamefont {Hao},
  \citenamefont {Ceder},\ and\ \citenamefont
  {McCloskey}}]{Huang2023_Mn_O_redox}%
  \BibitemOpen
  \bibfield  {author} {\bibinfo {author} {\bibfnamefont {T.~Y.}\ \bibnamefont
  {Huang}}, \bibinfo {author} {\bibfnamefont {Z.}~\bibnamefont {Cai}}, \bibinfo
  {author} {\bibfnamefont {M.~J.}\ \bibnamefont {Crafton}}, \bibinfo {author}
  {\bibfnamefont {L.~A.}\ \bibnamefont {Kaufman}}, \bibinfo {author}
  {\bibfnamefont {Z.~M.}\ \bibnamefont {Konz}}, \bibinfo {author}
  {\bibfnamefont {H.~K.}\ \bibnamefont {Bergstrom}}, \bibinfo {author}
  {\bibfnamefont {E.~A.}\ \bibnamefont {Kedzie}}, \bibinfo {author}
  {\bibfnamefont {H.~M.}\ \bibnamefont {Hao}}, \bibinfo {author} {\bibfnamefont
  {G.}~\bibnamefont {Ceder}},\ and\ \bibinfo {author} {\bibfnamefont {B.~D.}\
  \bibnamefont {McCloskey}},\ }\bibfield  {title} {\bibinfo {title}
  {{Quantitative Decoupling of Oxygen-Redox and Manganese-Redox Voltage
  Hysteresis in a Cation-Disordered Rock Salt Cathode}},\ }\href
  {https://doi.org/10.1002/aenm.202300241} {\bibfield  {journal} {\bibinfo
  {journal} {Advanced Energy Materials}\ }\textbf {\bibinfo {volume}
  {2300241}},\ \bibinfo {pages} {1} (\bibinfo {year}
  {2023}{\natexlab{b}})}\BibitemShut {NoStop}%
\bibitem [{\citenamefont {Ceder}(1993)}]{Ceder1993}%
  \BibitemOpen
  \bibfield  {author} {\bibinfo {author} {\bibfnamefont {G.}~\bibnamefont
  {Ceder}},\ }\bibfield  {title} {\bibinfo {title} {{A derivation of the Ising
  model for the computation of phase diagrams}},\ }\href
  {https://doi.org/10.1016/0927-0256(93)90005-8} {\bibfield  {journal}
  {\bibinfo  {journal} {Computational Materials Science}\ }\textbf {\bibinfo
  {volume} {1}},\ \bibinfo {pages} {144} (\bibinfo {year} {1993})}\BibitemShut
  {NoStop}%
\bibitem [{\citenamefont {M{\'{e}}n{\'{e}}trier}\ \emph
  {et~al.}(1999)\citenamefont {M{\'{e}}n{\'{e}}trier}, \citenamefont
  {Saadoune}, \citenamefont {Levasseur},\ and\ \citenamefont
  {Delmas}}]{Menetrier1999_MIT_LiCoO2}%
  \BibitemOpen
  \bibfield  {author} {\bibinfo {author} {\bibfnamefont {M.}~\bibnamefont
  {M{\'{e}}n{\'{e}}trier}}, \bibinfo {author} {\bibfnamefont {I.}~\bibnamefont
  {Saadoune}}, \bibinfo {author} {\bibfnamefont {S.}~\bibnamefont
  {Levasseur}},\ and\ \bibinfo {author} {\bibfnamefont {C.}~\bibnamefont
  {Delmas}},\ }\bibfield  {title} {\bibinfo {title} {{The insulator-metal
  transition upon lithium deintercalation from LiCoO$_2$: electronic properties
  and $^7$Li NMR study}},\ }\href {https://doi.org/10.1039/a900016j} {\bibfield
   {journal} {\bibinfo  {journal} {Journal of Materials Chemistry}\ }\textbf
  {\bibinfo {volume} {9}},\ \bibinfo {pages} {1135} (\bibinfo {year}
  {1999})}\BibitemShut {NoStop}%
\bibitem [{\citenamefont {Zunger}\ \emph {et~al.}(1990)\citenamefont {Zunger},
  \citenamefont {Wei}, \citenamefont {Ferreira},\ and\ \citenamefont
  {Bernard}}]{Zunger1990_SQS}%
  \BibitemOpen
  \bibfield  {author} {\bibinfo {author} {\bibfnamefont {A.}~\bibnamefont
  {Zunger}}, \bibinfo {author} {\bibfnamefont {S.-H.}\ \bibnamefont {Wei}},
  \bibinfo {author} {\bibfnamefont {L.~G.}\ \bibnamefont {Ferreira}},\ and\
  \bibinfo {author} {\bibfnamefont {J.~E.}\ \bibnamefont {Bernard}},\
  }\bibfield  {title} {\bibinfo {title} {{Special quasirandom structures}},\
  }\href {https://doi.org/10.1103/PhysRevLett.65.353} {\bibfield  {journal}
  {\bibinfo  {journal} {Physical Review Letters}\ }\textbf {\bibinfo {volume}
  {65}},\ \bibinfo {pages} {353} (\bibinfo {year} {1990})}\BibitemShut
  {NoStop}%
\bibitem [{\citenamefont {Huang}\ \emph {et~al.}(2021)\citenamefont {Huang},
  \citenamefont {Zhong}, \citenamefont {Ha}, \citenamefont {Kwon},
  \citenamefont {Crafton}, \citenamefont {Tian}, \citenamefont
  {Balasubramanian}, \citenamefont {McCloskey}, \citenamefont {Yang},\ and\
  \citenamefont {Ceder}}]{Huang2021_Cr}%
  \BibitemOpen
  \bibfield  {author} {\bibinfo {author} {\bibfnamefont {J.}~\bibnamefont
  {Huang}}, \bibinfo {author} {\bibfnamefont {P.}~\bibnamefont {Zhong}},
  \bibinfo {author} {\bibfnamefont {Y.}~\bibnamefont {Ha}}, \bibinfo {author}
  {\bibfnamefont {D.-h.}\ \bibnamefont {Kwon}}, \bibinfo {author}
  {\bibfnamefont {M.~J.}\ \bibnamefont {Crafton}}, \bibinfo {author}
  {\bibfnamefont {Y.}~\bibnamefont {Tian}}, \bibinfo {author} {\bibfnamefont
  {M.}~\bibnamefont {Balasubramanian}}, \bibinfo {author} {\bibfnamefont
  {B.~D.}\ \bibnamefont {McCloskey}}, \bibinfo {author} {\bibfnamefont
  {W.}~\bibnamefont {Yang}},\ and\ \bibinfo {author} {\bibfnamefont
  {G.}~\bibnamefont {Ceder}},\ }\bibfield  {title} {\bibinfo {title}
  {{Non-topotactic reactions enable high rate capability in Li-rich cathode
  materials}},\ }\href {https://doi.org/10.1038/s41560-021-00817-6} {\bibfield
  {journal} {\bibinfo  {journal} {Nature Energy}\ }\textbf {\bibinfo {volume}
  {6}},\ \bibinfo {pages} {706} (\bibinfo {year} {2021})}\BibitemShut {NoStop}%
\bibitem [{\citenamefont {Lun}\ \emph {et~al.}(2021)\citenamefont {Lun},
  \citenamefont {Ouyang}, \citenamefont {Kwon}, \citenamefont {Ha},
  \citenamefont {Foley}, \citenamefont {Huang}, \citenamefont {Cai},
  \citenamefont {Kim}, \citenamefont {Balasubramanian}, \citenamefont {Sun},
  \citenamefont {Huang}, \citenamefont {Tian}, \citenamefont {Kim},
  \citenamefont {McCloskey}, \citenamefont {Yang}, \citenamefont
  {Cl{\'{e}}ment}, \citenamefont {Ji},\ and\ \citenamefont
  {Ceder}}]{Lun2020_high_entropy}%
  \BibitemOpen
  \bibfield  {author} {\bibinfo {author} {\bibfnamefont {Z.}~\bibnamefont
  {Lun}}, \bibinfo {author} {\bibfnamefont {B.}~\bibnamefont {Ouyang}},
  \bibinfo {author} {\bibfnamefont {D.-h.}\ \bibnamefont {Kwon}}, \bibinfo
  {author} {\bibfnamefont {Y.}~\bibnamefont {Ha}}, \bibinfo {author}
  {\bibfnamefont {E.~E.}\ \bibnamefont {Foley}}, \bibinfo {author}
  {\bibfnamefont {T.-Y.}\ \bibnamefont {Huang}}, \bibinfo {author}
  {\bibfnamefont {Z.}~\bibnamefont {Cai}}, \bibinfo {author} {\bibfnamefont
  {H.}~\bibnamefont {Kim}}, \bibinfo {author} {\bibfnamefont {M.}~\bibnamefont
  {Balasubramanian}}, \bibinfo {author} {\bibfnamefont {Y.}~\bibnamefont
  {Sun}}, \bibinfo {author} {\bibfnamefont {J.}~\bibnamefont {Huang}}, \bibinfo
  {author} {\bibfnamefont {Y.}~\bibnamefont {Tian}}, \bibinfo {author}
  {\bibfnamefont {H.}~\bibnamefont {Kim}}, \bibinfo {author} {\bibfnamefont
  {B.~D.}\ \bibnamefont {McCloskey}}, \bibinfo {author} {\bibfnamefont
  {W.}~\bibnamefont {Yang}}, \bibinfo {author} {\bibfnamefont {R.~J.}\
  \bibnamefont {Cl{\'{e}}ment}}, \bibinfo {author} {\bibfnamefont
  {H.}~\bibnamefont {Ji}},\ and\ \bibinfo {author} {\bibfnamefont
  {G.}~\bibnamefont {Ceder}},\ }\bibfield  {title} {\bibinfo {title}
  {{Cation-disordered rocksalt-type high-entropy cathodes for Li-ion
  batteries}},\ }\href {https://doi.org/10.1038/s41563-020-00816-0} {\bibfield
  {journal} {\bibinfo  {journal} {Nature Materials}\ }\textbf {\bibinfo
  {volume} {20}},\ \bibinfo {pages} {214} (\bibinfo {year} {2021})}\BibitemShut
  {NoStop}%
\bibitem [{\citenamefont {Nakajima}\ and\ \citenamefont
  {Yabuuchi}(2017)}]{Nakajima2017_V_DRX}%
  \BibitemOpen
  \bibfield  {author} {\bibinfo {author} {\bibfnamefont {M.}~\bibnamefont
  {Nakajima}}\ and\ \bibinfo {author} {\bibfnamefont {N.}~\bibnamefont
  {Yabuuchi}},\ }\bibfield  {title} {\bibinfo {title} {{Lithium-Excess
  Cation-Disordered Rocksalt-Type Oxide with Nanoscale Phase Segregation:
  Li$_{1.25}$Nb$_{0.25}$V$_{0.5}$O$_{2}$}},\ }\href
  {https://doi.org/10.1021/acs.chemmater.7b02343} {\bibfield  {journal}
  {\bibinfo  {journal} {Chemistry of Materials}\ }\textbf {\bibinfo {volume}
  {29}},\ \bibinfo {pages} {6927} (\bibinfo {year} {2017})}\BibitemShut
  {NoStop}%
\bibitem [{\citenamefont {Lebens-Higgins}\ \emph {et~al.}(2021)\citenamefont
  {Lebens-Higgins}, \citenamefont {Chung}, \citenamefont {Temprano},
  \citenamefont {Zuba}, \citenamefont {Wu}, \citenamefont {Rana}, \citenamefont
  {Mejia}, \citenamefont {Jones}, \citenamefont {Wang}, \citenamefont {Grey}
  \emph {et~al.}}]{lebens2021electrochemical}%
  \BibitemOpen
  \bibfield  {author} {\bibinfo {author} {\bibfnamefont {Z.}~\bibnamefont
  {Lebens-Higgins}}, \bibinfo {author} {\bibfnamefont {H.}~\bibnamefont
  {Chung}}, \bibinfo {author} {\bibfnamefont {I.}~\bibnamefont {Temprano}},
  \bibinfo {author} {\bibfnamefont {M.}~\bibnamefont {Zuba}}, \bibinfo {author}
  {\bibfnamefont {J.}~\bibnamefont {Wu}}, \bibinfo {author} {\bibfnamefont
  {J.}~\bibnamefont {Rana}}, \bibinfo {author} {\bibfnamefont {C.}~\bibnamefont
  {Mejia}}, \bibinfo {author} {\bibfnamefont {M.~A.}\ \bibnamefont {Jones}},
  \bibinfo {author} {\bibfnamefont {L.}~\bibnamefont {Wang}}, \bibinfo {author}
  {\bibfnamefont {C.~P.}\ \bibnamefont {Grey}}, \emph {et~al.},\ }\bibfield
  {title} {\bibinfo {title} {{Electrochemical Utilization of Iron IV in the
  Li$_{1.3}$Fe$_{0.4}$Nb$_{0.3}$O$_2$ Disordered Rocksalt Cathode}},\ }\href
  {https://chemistry-europe.onlinelibrary.wiley.com/doi/abs/10.1002/batt.202000318}
  {\bibfield  {journal} {\bibinfo  {journal} {Batteries \& Supercaps}\ }\textbf
  {\bibinfo {volume} {4}},\ \bibinfo {pages} {771} (\bibinfo {year}
  {2021})}\BibitemShut {NoStop}%
\bibitem [{\citenamefont {Weinan}(2020)}]{weinan2020machine}%
  \BibitemOpen
  \bibfield  {author} {\bibinfo {author} {\bibfnamefont {E.}~\bibnamefont
  {Weinan}},\ }\bibfield  {title} {\bibinfo {title} {Machine learning and
  computational mathematics},\ }\href {https://arxiv.org/abs/2009.14596}
  {\bibfield  {journal} {\bibinfo  {journal} {arXiv preprint arXiv:2009.14596}\
  } (\bibinfo {year} {2020})}\BibitemShut {NoStop}%
\bibitem [{\citenamefont {Xie}\ \emph {et~al.}(2022)\citenamefont {Xie},
  \citenamefont {Zhou}, \citenamefont {Luan},\ and\ \citenamefont
  {Jiang}}]{Xie2022_MLFF_CE}%
  \BibitemOpen
  \bibfield  {author} {\bibinfo {author} {\bibfnamefont {J.~Z.}\ \bibnamefont
  {Xie}}, \bibinfo {author} {\bibfnamefont {X.~Y.}\ \bibnamefont {Zhou}},
  \bibinfo {author} {\bibfnamefont {D.}~\bibnamefont {Luan}},\ and\ \bibinfo
  {author} {\bibfnamefont {H.}~\bibnamefont {Jiang}},\ }\bibfield  {title}
  {\bibinfo {title} {{Machine Learning Force Field Aided Cluster Expansion
  Approach to Configurationally Disordered Materials: Critical Assessment of
  Training Set Selection and Size Convergence}},\ }\href
  {https://doi.org/10.1021/acs.jctc.2c00017} {\bibfield  {journal} {\bibinfo
  {journal} {Journal of Chemical Theory and Computation}\ }\textbf {\bibinfo
  {volume} {18}},\ \bibinfo {pages} {3795} (\bibinfo {year}
  {2022})}\BibitemShut {NoStop}%
\bibitem [{\citenamefont {Chen}\ and\ \citenamefont
  {Ong}(2022)}]{chen2022_m3gnet}%
  \BibitemOpen
  \bibfield  {author} {\bibinfo {author} {\bibfnamefont {C.}~\bibnamefont
  {Chen}}\ and\ \bibinfo {author} {\bibfnamefont {S.~P.}\ \bibnamefont {Ong}},\
  }\bibfield  {title} {\bibinfo {title} {{A universal graph deep learning
  interatomic potential for the periodic table}},\ }\href
  {https://doi.org/10.1038/s43588-022-00349-3} {\bibfield  {journal} {\bibinfo
  {journal} {Nature Computational Science}\ }\textbf {\bibinfo {volume} {2}},\
  \bibinfo {pages} {718} (\bibinfo {year} {2022})}\BibitemShut {NoStop}%
\bibitem [{\citenamefont {Deng}\ \emph {et~al.}(2023)\citenamefont {Deng},
  \citenamefont {Zhong}, \citenamefont {Jun}, \citenamefont {Riebesell},
  \citenamefont {Han}, \citenamefont {Bartel},\ and\ \citenamefont
  {Ceder}}]{deng2023chgnet}%
  \BibitemOpen
  \bibfield  {author} {\bibinfo {author} {\bibfnamefont {B.}~\bibnamefont
  {Deng}}, \bibinfo {author} {\bibfnamefont {P.}~\bibnamefont {Zhong}},
  \bibinfo {author} {\bibfnamefont {K.}~\bibnamefont {Jun}}, \bibinfo {author}
  {\bibfnamefont {J.}~\bibnamefont {Riebesell}}, \bibinfo {author}
  {\bibfnamefont {K.}~\bibnamefont {Han}}, \bibinfo {author} {\bibfnamefont
  {C.~J.}\ \bibnamefont {Bartel}},\ and\ \bibinfo {author} {\bibfnamefont
  {G.}~\bibnamefont {Ceder}},\ }\bibfield  {title} {\bibinfo {title} {{CHGNet:
  Pretrained universal neural network potential for charge-informed atomistic
  modeling}},\ }\href {https://arxiv.org/abs/2302.14231} {\bibfield  {journal}
  {\bibinfo  {journal} {arXiv preprint arXiv:2302.14231}\ } (\bibinfo {year}
  {2023})}\BibitemShut {NoStop}%
\bibitem [{\citenamefont {Takamoto}\ \emph {et~al.}(2022)\citenamefont
  {Takamoto}, \citenamefont {Shinagawa}, \citenamefont {Motoki}, \citenamefont
  {Nakago}, \citenamefont {Li}, \citenamefont {Kurata}, \citenamefont
  {Watanabe}, \citenamefont {Yayama}, \citenamefont {Iriguchi}, \citenamefont
  {Asano}, \citenamefont {Onodera}, \citenamefont {Ishii}, \citenamefont
  {Kudo}, \citenamefont {Ono}, \citenamefont {Sawada}, \citenamefont
  {Ishitani}, \citenamefont {Ong}, \citenamefont {Yamaguchi}, \citenamefont
  {Kataoka}, \citenamefont {Hayashi}, \citenamefont {Charoenphakdee},\ and\
  \citenamefont {Ibuka}}]{Takamoto2022_PFP}%
  \BibitemOpen
  \bibfield  {author} {\bibinfo {author} {\bibfnamefont {S.}~\bibnamefont
  {Takamoto}}, \bibinfo {author} {\bibfnamefont {C.}~\bibnamefont {Shinagawa}},
  \bibinfo {author} {\bibfnamefont {D.}~\bibnamefont {Motoki}}, \bibinfo
  {author} {\bibfnamefont {K.}~\bibnamefont {Nakago}}, \bibinfo {author}
  {\bibfnamefont {W.}~\bibnamefont {Li}}, \bibinfo {author} {\bibfnamefont
  {I.}~\bibnamefont {Kurata}}, \bibinfo {author} {\bibfnamefont
  {T.}~\bibnamefont {Watanabe}}, \bibinfo {author} {\bibfnamefont
  {Y.}~\bibnamefont {Yayama}}, \bibinfo {author} {\bibfnamefont
  {H.}~\bibnamefont {Iriguchi}}, \bibinfo {author} {\bibfnamefont
  {Y.}~\bibnamefont {Asano}}, \bibinfo {author} {\bibfnamefont
  {T.}~\bibnamefont {Onodera}}, \bibinfo {author} {\bibfnamefont
  {T.}~\bibnamefont {Ishii}}, \bibinfo {author} {\bibfnamefont
  {T.}~\bibnamefont {Kudo}}, \bibinfo {author} {\bibfnamefont {H.}~\bibnamefont
  {Ono}}, \bibinfo {author} {\bibfnamefont {R.}~\bibnamefont {Sawada}},
  \bibinfo {author} {\bibfnamefont {R.}~\bibnamefont {Ishitani}}, \bibinfo
  {author} {\bibfnamefont {M.}~\bibnamefont {Ong}}, \bibinfo {author}
  {\bibfnamefont {T.}~\bibnamefont {Yamaguchi}}, \bibinfo {author}
  {\bibfnamefont {T.}~\bibnamefont {Kataoka}}, \bibinfo {author} {\bibfnamefont
  {A.}~\bibnamefont {Hayashi}}, \bibinfo {author} {\bibfnamefont
  {N.}~\bibnamefont {Charoenphakdee}},\ and\ \bibinfo {author} {\bibfnamefont
  {T.}~\bibnamefont {Ibuka}},\ }\bibfield  {title} {\bibinfo {title} {{Towards
  universal neural network potential for material discovery applicable to
  arbitrary combination of 45 elements}},\ }\href
  {https://doi.org/10.1038/s41467-022-30687-9} {\bibfield  {journal} {\bibinfo
  {journal} {Nature Communications}\ }\textbf {\bibinfo {volume} {13}},\
  \bibinfo {pages} {1} (\bibinfo {year} {2022})}\BibitemShut {NoStop}%
\bibitem [{\citenamefont {Kresse}\ and\ \citenamefont
  {Furthm{\"{u}}ller}(1996)}]{kresse1996VASP}%
  \BibitemOpen
  \bibfield  {author} {\bibinfo {author} {\bibfnamefont {G.}~\bibnamefont
  {Kresse}}\ and\ \bibinfo {author} {\bibfnamefont {J.}~\bibnamefont
  {Furthm{\"{u}}ller}},\ }\bibfield  {title} {\bibinfo {title} {{Efficiency of
  ab-initio total energy calculations for metals and semiconductors using a
  plane-wave basis set}},\ }\href
  {https://doi.org/10.1016/0927-0256(96)00008-0} {\bibfield  {journal}
  {\bibinfo  {journal} {Computational Materials Science}\ }\textbf {\bibinfo
  {volume} {6}},\ \bibinfo {pages} {15} (\bibinfo {year} {1996})}\BibitemShut
  {NoStop}%
\bibitem [{\citenamefont {Kresse}\ and\ \citenamefont
  {Joubert}(1999)}]{kresse1999PAW}%
  \BibitemOpen
  \bibfield  {author} {\bibinfo {author} {\bibfnamefont {G.}~\bibnamefont
  {Kresse}}\ and\ \bibinfo {author} {\bibfnamefont {D.}~\bibnamefont
  {Joubert}},\ }\bibfield  {title} {\bibinfo {title} {{From ultrasoft
  pseudopotentials to the projector augmented-wave method}},\ }\href
  {https://doi.org/10.1103/PhysRevB.59.1758} {\bibfield  {journal} {\bibinfo
  {journal} {Physical Review B}\ }\textbf {\bibinfo {volume} {59}},\ \bibinfo
  {pages} {1758} (\bibinfo {year} {1999})}\BibitemShut {NoStop}%
\bibitem [{\citenamefont {Furness}\ \emph {et~al.}(2020)\citenamefont
  {Furness}, \citenamefont {Kaplan}, \citenamefont {Ning}, \citenamefont
  {Perdew},\ and\ \citenamefont {Sun}}]{furness2020r2SCAN}%
  \BibitemOpen
  \bibfield  {author} {\bibinfo {author} {\bibfnamefont {J.~W.}\ \bibnamefont
  {Furness}}, \bibinfo {author} {\bibfnamefont {A.~D.}\ \bibnamefont {Kaplan}},
  \bibinfo {author} {\bibfnamefont {J.}~\bibnamefont {Ning}}, \bibinfo {author}
  {\bibfnamefont {J.~P.}\ \bibnamefont {Perdew}},\ and\ \bibinfo {author}
  {\bibfnamefont {J.}~\bibnamefont {Sun}},\ }\bibfield  {title} {\bibinfo
  {title} {{Accurate and Numerically Efficient r$^2$SCAN Meta-Generalized
  Gradient Approximation}},\ }\href
  {https://doi.org/10.1021/acs.jpclett.0c02405} {\bibfield  {journal} {\bibinfo
   {journal} {The Journal of Physical Chemistry Letters}\ }\textbf {\bibinfo
  {volume} {11}},\ \bibinfo {pages} {8208} (\bibinfo {year} {2020})},\ \Eprint
  {https://arxiv.org/abs/2008.03374} {2008.03374} \BibitemShut {NoStop}%
\bibitem [{\citenamefont {Zhang}\ \emph {et~al.}(2018)\citenamefont {Zhang},
  \citenamefont {Kitchaev}, \citenamefont {Yang}, \citenamefont {Chen},
  \citenamefont {Dacek}, \citenamefont {Sarmiento-P{\'e}rez}, \citenamefont
  {Marques}, \citenamefont {Peng}, \citenamefont {Ceder}, \citenamefont
  {Perdew} \emph {et~al.}}]{zhang2018_npjSCAN}%
  \BibitemOpen
  \bibfield  {author} {\bibinfo {author} {\bibfnamefont {Y.}~\bibnamefont
  {Zhang}}, \bibinfo {author} {\bibfnamefont {D.~A.}\ \bibnamefont {Kitchaev}},
  \bibinfo {author} {\bibfnamefont {J.}~\bibnamefont {Yang}}, \bibinfo {author}
  {\bibfnamefont {T.}~\bibnamefont {Chen}}, \bibinfo {author} {\bibfnamefont
  {S.~T.}\ \bibnamefont {Dacek}}, \bibinfo {author} {\bibfnamefont {R.~A.}\
  \bibnamefont {Sarmiento-P{\'e}rez}}, \bibinfo {author} {\bibfnamefont
  {M.~A.}\ \bibnamefont {Marques}}, \bibinfo {author} {\bibfnamefont
  {H.}~\bibnamefont {Peng}}, \bibinfo {author} {\bibfnamefont {G.}~\bibnamefont
  {Ceder}}, \bibinfo {author} {\bibfnamefont {J.~P.}\ \bibnamefont {Perdew}},
  \emph {et~al.},\ }\bibfield  {title} {\bibinfo {title} {Efficient
  first-principles prediction of solid stability: Towards chemical accuracy},\
  }\href {https://www.nature.com/articles/s41524-018-0065-z} {\bibfield
  {journal} {\bibinfo  {journal} {npj Computational Materials}\ }\textbf
  {\bibinfo {volume} {4}},\ \bibinfo {pages} {1} (\bibinfo {year}
  {2018})}\BibitemShut {NoStop}%
\bibitem [{\citenamefont {Kingsbury}\ \emph {et~al.}(2022)\citenamefont
  {Kingsbury}, \citenamefont {Gupta}, \citenamefont {Bartel}, \citenamefont
  {Munro}, \citenamefont {Dwaraknath}, \citenamefont {Horton},\ and\
  \citenamefont {Persson}}]{kingsbury2022r2SCAN_PRM}%
  \BibitemOpen
  \bibfield  {author} {\bibinfo {author} {\bibfnamefont {R.}~\bibnamefont
  {Kingsbury}}, \bibinfo {author} {\bibfnamefont {A.~S.}\ \bibnamefont
  {Gupta}}, \bibinfo {author} {\bibfnamefont {C.~J.}\ \bibnamefont {Bartel}},
  \bibinfo {author} {\bibfnamefont {J.~M.}\ \bibnamefont {Munro}}, \bibinfo
  {author} {\bibfnamefont {S.}~\bibnamefont {Dwaraknath}}, \bibinfo {author}
  {\bibfnamefont {M.}~\bibnamefont {Horton}},\ and\ \bibinfo {author}
  {\bibfnamefont {K.~A.}\ \bibnamefont {Persson}},\ }\bibfield  {title}
  {\bibinfo {title} {{Performance comparison of r$^2$SCAN and SCAN metaGGA
  density functionals for solid materials via an automated, high-throughput
  computational workflow}},\ }\href
  {https://journals.aps.org/prmaterials/abstract/10.1103/PhysRevMaterials.6.013801}
  {\bibfield  {journal} {\bibinfo  {journal} {Physical Review Materials}\
  }\textbf {\bibinfo {volume} {6}},\ \bibinfo {pages} {013801} (\bibinfo {year}
  {2022})}\BibitemShut {NoStop}%
\end{thebibliography}%
\end{document}